\documentclass[journal,a4paper]{IEEEtran}
\usepackage[utf8]{inputenc}
\usepackage{amsmath}
\usepackage{mathtools}
\usepackage{graphicx}
\usepackage{float}
\usepackage{pgfplots}
\usepackage{tikz}
\usetikzlibrary{plotmarks}
\usetikzlibrary{external}
\usetikzlibrary{arrows}
\usetikzlibrary{shapes,decorations}
\usetikzlibrary{colorbrewer}
\usepackage{caption}
\usepackage{cite}
\pgfplotsset{compat=1.14}
\usepackage{xcolor}
\usepackage{hyperref}
\usetikzlibrary{shapes}
\usepackage{pgfplotstable}
\usetikzlibrary{plotmarks}
\usetikzlibrary{external}
\usetikzlibrary{arrows}
\usetikzlibrary{shapes,decorations}
\usetikzlibrary{colorbrewer}
\usepackage{caption}
\pgfplotsset{compat=1.14}
\usepackage[utf8]{inputenc}
\newcommand{\rom}[1]{\uppercase\expandafter{\romannumeral #1\relax}}
\newcommand{\indB}{\left[\text{dB}\right]}
\usepackage{amssymb}

\begin{document}

\title{A Closed-Form Approximation of the Gaussian Noise Model in the Presence of Inter-Channel Stimulated Raman Scattering}
%
%
%

\author{Daniel~Semrau,~\IEEEmembership{Student Member,~IEEE,} Robert~I.~Killey,~\IEEEmembership{Senior Member,~IEEE,} and~Polina~Bayvel,~\IEEEmembership{Fellow,~IEEE,~Fellow,~OSA}
\thanks{This work was supported by a UK EPSRC programme grant TRANSNET (EP/R035342/1) and a Doctoral Training Partnership (DTP) studentship for Daniel Semrau.}
\thanks{D. Semrau, Robert I. Killey and P. Bayvel are with the Optical Networks Group, University College London, London
WC1E 7JE, U.K. (e-mail: \{uceedfs; r.killey; p.bayvel\}@ucl.ac.uk.)}
}

\maketitle

\markboth{\today}%
{}

\begin{abstract}
An accurate, closed-form expression evaluating the nonlinear interference (NLI) power in coherent optical transmission systems in the presence of inter-channel stimulated Raman scattering (ISRS) is derived. The analytical result enables a rapid estimate of the signal-to-noise ratio (SNR) and avoids the need for integral evaluations and split-step simulations. The formula also provides new insight into the underlying parameter dependence of ISRS on the NLI. The proposed result is applicable for dispersion unmanaged, ultra-wideband transmission systems that use optical bandwidths of up to 15~THz. 
\par 
\ 
The accuracy of the closed-form expression is compared to numerical integration of the ISRS Gaussian Noise model and split-step simulations in a point-to-point transmission, as well as in a mesh optical network scenario. 
\end{abstract}


\IEEEpeerreviewmaketitle

\section{Introduction}

\IEEEPARstart{A}{nalytical} models to estimate nonlinear interference (NLI) are key for rapid and efficient system design \cite{Hasegawa_2017_ofd}, achievable rate estimations of point-to-point links \cite{Semrau_2016_air,Bosco_2011_aro,Shevchenko_2016_air} and physical layer aware network optimization. The latter is essential for optical network abstraction and virtualization leading to optimal and intelligent techniques to maximize optical network capacity \cite{Anagnostopoulos_2007_pli}.
\par 
\ 
Most approaches analytically solve the nonlinear Schr\"odinger equation using first-order perturbation theory with respect to the Kerr nonlinearity. The resulting integral expressions offer a significant reduction in computational complexity with minor inaccuracies compared to split-step simulations and experiments \cite{Nespola_2014_gvo,Nespola_2015_evo,Galdino_2016_edo,Saavedra_2017_eio,Saavedra_2017_eao,Saavedra_2018_isr}.
\par 
\ 
Particularly, the Gaussian Noise (GN) model offers reasonable accuracy while exhibiting a moderate computational complexity \cite{Tang_2002_tcc,Poggiolini_2012_tgm}. Numerical integration of the GN model to obtain the nonlinear interference has a typical computation time of a few minutes per WDM channel \cite{Semrau_2018_tig,Cantono_2018_oti}. For ultra-wideband signals that consist of 200 WDM channels or more, the computation time quickly amounts to a few hours to obtain the NLI distribution across the entire optical signal. 
\begin{figure}[h]
\centering
\newcommand\W{0.5} 

\newcommand\X{0} 
\newcommand\XX{2.1} 
\newcommand\XXX{3.9} 
\newcommand\XXXX{5.45} 
\newcommand\XXXXX{6.9} 
\newcommand\Y{0} 
\newcommand\YY{1.1} 
\newcommand\YYY{2.2 } 
\newcommand\YYYY{3.7} 
\newcommand\YYYYY{5.2} 
\newcommand\YYYYYY{5.8} 
\includegraphics[]{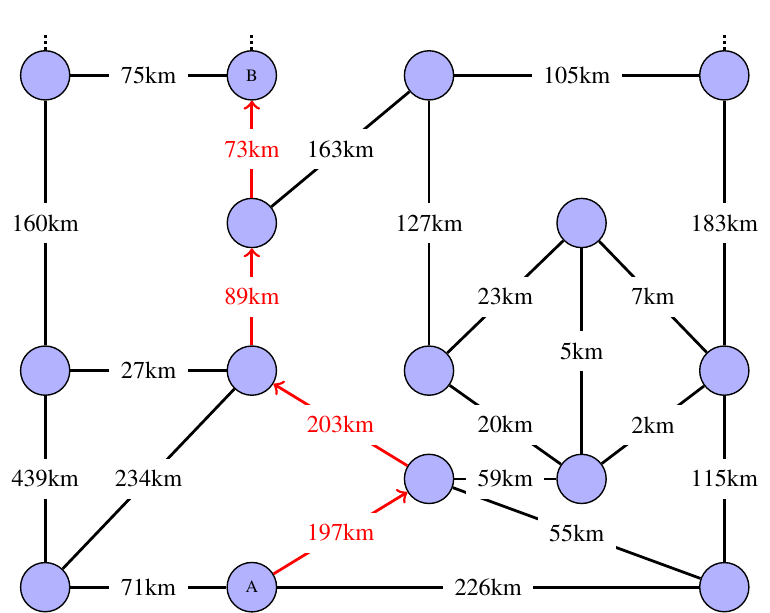}
\caption{A section from the British Telecommunications (BT) 20+2 topology of the United Kingdom (UK) core network \cite{Ives_2014_qti}. In section \ref{sec:mesh_network}, the nonlinear interference of the example light path (A-B) is modeled.}
\label{fig:network_topology}
\end{figure}
\par 
\ 
For some applications, such time frames are not acceptable and closed-form approximations, that yield performance estimations in picoseconds, are preferable. Such applications include e.g. physical layer aware network optimization and network performance estimation. Additionally, closed-form approximations offer a unique insight into the underlying parameter dependencies and provide useful design and scaling rules. 
\par 
\ 
An optical mesh network consists of a vast amount of point-to-point transmissions; two between two adjacent reconfigurable optical add-drop multiplexer (ROADM). As an example, the British Telecommunications (BT) 20+2 topology of the United Kingdom core network \cite{Ives_2014_qti} exhibits 34 bidirectional network edges. A section of the network is shown in Fig. \ref{fig:network_topology}. For ultra-wideband transmission, assuming a WDM grid spacing of 50~GHz (ITU grid) over the entire C+L band (10~THz), a full performance estimation of a \textit{single} network state requires up to $\frac{10\text{THz}}{50\text{GHz}}\cdot 2\cdot 34=13600$ NLI function evaluations. Furthermore, network operators often seek for optimum channel allocations and optimum launch powers for new incoming demands, requiring performance computations of numerous potential network states. It is evident that the problem is intractable using numerical integrations. However, using closed-form approximations, the performance estimation of a given network state can be reduced to only a few microseconds. 
\par 
\ 
Closed-form approximations, that predict the nonlinear interference power in coherent transmission systems, were derived for lossless fibers \cite{Splett_1993_utc,Bosco_2011_aro} and lossy fibers using lumped amplification \cite{Louchet_2003_amf,Chen_2010_cef,Poggiolini_2012_tgm,Savory_2013_aft,Johannisson_2014_mon,Poggiolini_2015_asa} as well as for distributed Raman amplified links \cite{Semrau_2017_ace}. 
\par 
\ 
However, all aforementioned formulas are not applicable for optical bandwidths beyond C-band (5~THz) where inter-channel stimulated Raman scattering (ISRS) becomes significant. ISRS is a non-parametric nonlinear effect that effectively amplifies low frequency components at the expense of high frequency components within the same optical signal. This significantly alters the NLI distribution across the received spectrum \cite{Semrau_17_ard}.
\par 
\ 
Recently, we introduced the ISRS GN model \cite{Semrau_2018_tgn,Semrau_2018_tig}, which is a GN model that rigorously accounts for ISRS and is, therefore, suitable for optical bandwidths beyond C-band. Comparable models were published in \cite{Semrau_17_ard,Roberts_17_cpo,Cantono_2018_mti,Cantono_2018_oti}. The ISRS GN model relies on numerically solving an integral of at least three dimensions and an approximation in closed-form has yet to be reported.

\par 
In this paper, a closed-form approximation of the ISRS GN model is presented which accurately accounts for the impact of ISRS on the nonlinear interference power. The proposed formula is applicable to dispersion unmanaged ultra-wideband transmission systems with optical bandwidths of up to 15~THz. Additionally, the proposed formula accounts for the dispersion slope and variably loaded fiber spans, enabling performance estimations in optical mesh networks. The analytical result is extensively validated by split-step simulations. 
\par 
The paper is organized as follows: The ISRS GN model is briefly revised in Section \ref{sec:TheISRSGNmodelA}. Section \ref{sec:xpm_assumption} addresses the key steps in the derivation of the closed-form approximation and the result is presented in Section \ref{sec:closed-form}. In Section \ref{sec:numerical_validation}, the proposed formula is validated via split-step simulations and via the ISRS GN model in integral form. The validation is carried out in a point-to-point scenario in Section \ref{sec:p2p_links} and in a mesh optical network scenario in Section \ref{sec:mesh_network}. 
\par 
\section{The ISRS GN model in closed-form}
\label{sec:TheISRSGNmodel}
In this section, the proposed closed-form approximation of the ISRS GN model is presented. 
The main derivation steps are outlined and its key assumptions are addressed and discussed.
\subsection{The ISRS GN model}
\label{sec:TheISRSGNmodelA}
In the following the ISRS GN model, a Gaussian model that accounts for inter-channel stimulated Raman scattering, is briefly
revised. After coherent detection, electronic dispersion compensation and neglecting the impact of transceiver noise, the signal-to-noise ratio (SNR) of the channel of interest (COI) $i$ can be calculated as
\begin{equation}
\begin{split}
\label{eq:SNR}
\textnormal{SNR}_i \approx \frac{P_i}{P_\textnormal{ASE} + \eta_n P_i^3},
\end{split}
\end{equation}
where $P_i$ is the launch power of channel $i$ and $P_\textnormal{ASE}$ is the accumulated amplified spontaneous emission (ASE) noise originating from optical amplifiers. The nonlinear interference coefficient $\eta_n\left(f_i\right)$ after $n$ spans is dependent on the center frequency $f_i$ of the COI. When the impact of ISRS is negligible, it is convenient to write the total nonlinear interference power as $P_\text{NLI}=\eta_n P_i^3$, as the NLI coefficient is independent of the absolute launch power and the cubic law in \eqref{eq:SNR} holds within the first-order perturbation regime.
\par 
\ 
However, for bandwidths beyond C-band, where ISRS cannot be neglected, the NLI coefficient itself becomes a function of the total launch power $P_{\text{tot}}$ and its absolute launch power distribution $G_{\text{Tx}}\left(f\right)$. Similar to \cite[Eq. (9)]{Semrau_2018_tgn}, the NLI coefficient in the presence of ISRS is given by  
\begin{equation}
\begin{split}
&\eta_1\left(f_i\right) = \frac{B_i}{P_i^3}\frac{16}{27}\gamma^2\int df_1 \int df_2 \ G_{\text{Tx}}(f_1)G_{\text{Tx}}(f_2)\\
&\cdot G_{\text{Tx}}(f_1+f_2-f_i) \\
&\cdot \left| \int_0^L d\zeta \ \frac{P_{\text{tot}}e^{-\alpha \zeta-P_{\text{tot}}C_{\text{r}} L_{\text{eff}}(f_1+f_2-f_i)}}{\int G_{\text{Tx}}(\nu)e^{-P_{\text{tot}}C_{\text{r}} L_{\text{eff}}\nu} d\nu}e^{j\phi\left(f_1,f_2,f_i,\zeta\right)}\right|^2,
\label{eq:ISRS_GNmodel}
\end{split}
\end{equation}
where $B_i$ is the bandwidth of the COI $i$, $\gamma$ is the nonlinearity coefficient, $\alpha$ is the attenuation coefficient, $L_{\text{eff}}=\frac{1-e^{-\alpha \zeta}}{\alpha}$ (the $\zeta$ dependence is suppressed throughout this paper for notational brevity), $C_r$ is the slope of the linear regression of the normalized Raman gain spectrum and $\phi=-4\pi^2(f_1-f)(f_2-f)\left[\beta_2+\pi\beta_3(f_1+f_2)\right]\zeta$ is a phase mismatch term, with the group velocity dispersion (GVD) parameter $\beta_2$ and its linear slope $\beta_3$. Eq. \eqref{eq:ISRS_GNmodel} is valid for optical bandwidths of up to 15~THz as it assumed a linear Raman gain spectrum. The ISRS GN model in integral form for multi-span systems ca be found in \cite[Eq. (2)]{Semrau_2018_tig}. Eq. \eqref{eq:ISRS_GNmodel} is obtained from the result derived in \cite[Eq. (9)]{Semrau_2018_tgn} with the difference that \eqref{eq:ISRS_GNmodel} assumes a uniform NLI PSD over the channel bandwidth (in the literature sometimes referred to as the local white noise assumption). This assumption is required in order to avoid the integration of the NLI PSD over the receiver filter (i.e. the matched filter), which is analytically difficult for arbitrary filter shapes.
\par
\
The strength of ISRS for a given system can be assessed by calculating the net power transfer $\Delta \rho \left(z\right)$ between the outer channels of the WDM signal. This is the ISRS net gain of the lowest frequency channel added to the ISRS net loss of the highest frequency channel, which is given by \cite[Eq. (8)]{Zirngibl_1998_amo}
\begin{equation}
\begin{split}
\Delta \rho \left(z\right)\indB \triangleq 10\log_{10}\left[\Delta \rho \left(z\right)\right]= 4.3\cdot P_{\text{tot}}C_{\text{r}} L_{\text{eff}}B_{\text{tot}},
\label{eq:max_pwrspread}
\end{split}
\end{equation}
where $B_{\text{tot}}$ is the total optical bandwidth.
\subsection{The XPM assumption}
\label{sec:xpm_assumption}
\begin{figure}[h]
\centering

\includegraphics[]{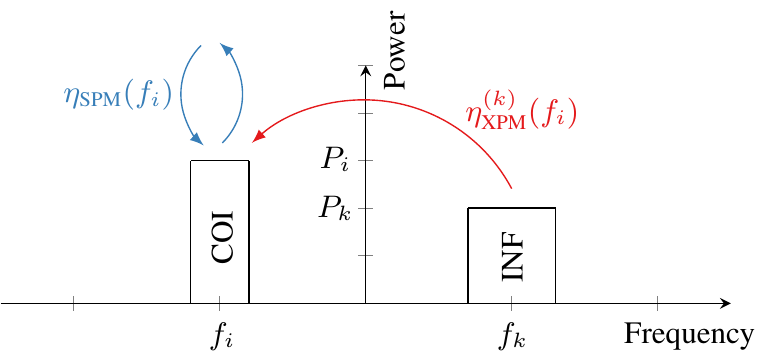}
\caption{Illustration of a transmitted spectrum $G_\text{Tx}\left(f\right)$ subject to the XPM assumption. Shown are the channel under test (COI) and a single interferer (INF) with arbitrary power levels, bandwidths and center frequencies. The total NLI is then obtained by summing over all interferers $k\in A_i$ as in \eqref{eq:XPM_set}.}
\label{fig:xpm_scheme}
\end{figure}

In the following, the key steps in deriving the proposed closed-form approximation of \eqref{eq:ISRS_GNmodel} are addressed and discussed. 
\par 
\         
We first evaluate the nonlinear perturbation of the COI $i$, caused by a single interferer (INF) $k$, which is denoted as $\eta_\text{XPM}^{\left(k\right)}(f_i)$. The special case where the NLI is caused by the COI itself (i.e. $k=i$), is denoted as SPM (sometimes referred to as SCI). The NLI contribution of all other interferer is denoted as XPM (sometimes referred to as XCI). An illustration of the SPM and XPM contribution of a COI and single INF is schematically shown in Fig. \ref{fig:xpm_scheme}. In more detail, the set of all XPM interferers, with respect to the COI $i$, is given as 
\begin{equation}
\begin{split}
A_i=\left\{k\in\mathbb{N} \ | \ 1 \leq k\leq N_\text{ch} \ \text{and} \  k\neq i \right\}.
\label{eq:XPM_set}
\end{split}
\end{equation}
In the literature, this assumption is often referred to as XPM assumption \cite{Mecozzi_2012_nsl,Secondini_2012_afc,Dar_2013_pon,Johannisson_2014_mon,Ives_2014_atp} and it neglects NLI contributions that are jointly generated by two interfering channels, which is denoted as FWM (sometimes referred to as MCI). However, this contribution is typically very small in high dispersive links, where high baud rates or channel spacings are used \cite{Carena_2014_emo,Zhang_2017_fae}.
\par 
\ 
Using the XPM assumption, the NLI coefficient can be written as
\begin{equation}
\begin{split}
&\eta_n\left(f_i\right)\approx\eta_\text{SPM}\left(f_i\right)n^{1+\epsilon}+\eta_\text{XPM}\left(f_i\right)n,
\label{eq:eta_total_XPM}
\end{split}
\end{equation}
where $\eta_\text{SPM}\left(f_i\right)$ is the SPM contribution and $\eta_\text{XPM}\left(f_i\right)$ is the total XPM contribution, which is the summation over all XPM interferer, as 
\begin{equation}
\begin{split}
&\eta_\text{XPM}\left(f_i\right)=\sum_{\forall k \in A_i}\eta_\text{XPM}^{\left(k\right)}(f_i),
\label{eq:XPM_eta}
\end{split}
\end{equation}
where $\eta_\text{XPM}^{\left(k\right)}(f_i)$ is the NLI (XPM) contribution of a single channel $k$ on channel $i$. 
\par 
\ 
The coherent accumulation along multiple fiber spans is included using the coherence factor $\epsilon$. The coherence factor can be obtained in closed-form \cite[Eq. (22)]{Poggiolini_2012_tgm}. The coherence factor is typically defined for the entire optical signal \cite{Poggiolini_2012_tgm,Semrau_2017_ace}. However in this work, only the SPM contribution is assumed to accumulate coherently and the coherence factor is redefined over the channel bandwidth $B_i$ \cite{Ives_2016_ati}. The XPM contribution is assumed to accumulate incoherently. The advantage of this approach is that the coherent accumulation is independent of the transmitted spectrum and only a function of the channel bandwidth. This significantly simplifies the modeling of NLI in optical mesh networks as fiber spans are variably loaded. The approach is consistent with the observations in \cite{Carena_2014_emo,Poggiolini_2015_asa}. 
\par 
\  
Additionally, it is assumed that the coherence factor itself is not altered by ISRS. In \cite{Semrau_2018_tgn}, it was shown that for standard single mode fiber (SMF) based spans and an ISRS power transfer of $\Delta \rho \left(z\right)\left[\text{dB}\right]=6.5$~dB over $10$~THz the coherence factor changes by up to $\Delta \epsilon=0.012$. Neglecting this effect, results in an approximation error of $\Delta \epsilon\cdot 10\log\left(10\right) =0.1$~dB in NLI after 10 spans and $\Delta \epsilon\cdot 10\log\left(50\right)=0.2$~dB after 50 spans. In this work, this effect is neglected and the coherent accumulation is assumed to be \textit{unaffected} by ISRS.
\par 
\ 
In the following, the NLI caused by an interferer $k$ on the COI $i$, is analytically evaluated. It is assumed that the channel of interest $i$ has normalized pulse shape $g_i(f-f_i)=\frac{1}{B_i}\Pi \left(\frac{f-f_i}{B_i}\right)$, launch power $P_i$, channel bandwidth $B_i$ and is centered around frequency $f_i$. The function $\Pi \left(x\right)$ denotes the rectangular function. The interfering channel has normalized pulse shape $g_k(f-f_k)=\frac{1}{B_k}\Pi \left(\frac{f-f_k}{B_k}\right)$, launch power $P_k$, bandwidth $B_k$ and is centered around frequency $f_k$. The transmitted spectrum is then given by
\begin{equation}
\begin{split}
G_\text{Tx}(f)=P_ig_i(f-f_i)+P_kg_k(f-f_i-\Delta f),
\label{eq:GTX}
\end{split}
\end{equation}
where $\Delta f=f_k-f_i$ is the frequency separation between COI and INF. An illustration of \eqref{eq:GTX} with the resulting nonlinear interactions on the COI is shown in Fig. \ref{fig:xpm_scheme}.
\par 
\ 
Substituting the transmitted spectrum \eqref{eq:GTX} in the ISRS GN model \eqref{eq:ISRS_GNmodel} yields six non-identical cross terms where only two are non-zero and contribute to the NLI of the COI. These two terms are the SPM and the XPM contribution. The XPM contribution is 
\begin{equation}
\begin{split}
&\eta_\text{XPM}^{\left(k\right)}(f_i) = \frac{32}{27}\frac{\gamma^2}{B^2_k}\left(\frac{P_k}{P_i}\right)^2\int_{-\frac{B_i}{2}}^{\frac{B_i}{2}} df_1 \int_{-\frac{B_k}{2}}^{\frac{B_k}{2}} df_2 \ \Pi\left(\frac{f_1+f_2}{B_k}\right) \\
&\cdot \left| \int_0^L d\zeta \ \frac{P_{\text{tot}}C_{\text{r}} L_{\text{eff}}B_\text{tot}}{2}\frac{e^{-\alpha \zeta-P_{\text{tot}}C_{\text{r}} L_{\text{eff}}(f_1+f_2+f_i+\Delta f)}}{\text{sinh}\left(\frac{P_{\text{tot}}C_{\text{r}} L_{\text{eff}}B_\text{tot}}{2}\right) }\right.\\
&\left.\cdot e^{j\phi\left(f_1+f_i,f_2+f_i+\Delta f,f_i,\zeta\right)}\right|^2,
\label{eq:XPM_integral}
\end{split}
\end{equation}
and the SPM contribution is 
\begin{equation}
\begin{split}
&\eta_\text{SPM}(f_i) = \frac{1}{2}\eta_\text{XPM}^{\left(i\right)}(f_i).
\label{eq:SPM_contribution}
\end{split}
\end{equation}
It should be noted that the quantities $P_\text{tot}$ and $B_\text{tot}$ refer to the launch power and bandwidth of the entire transmitted WDM signal and \textit{not} to the transmitted spectrum of a single COI and INF pair. 
\par 
\ 
ISRS alters the effective attenuation for a given WDM channel during propagation and hence introduces a change in the signal power profile within a fiber span. This change is a function of the total transmitted launch power, mathematically expressed as $P_{\text{tot}}e^{-P_{\text{tot}}C_{\text{r}} L_{\text{eff}}(f_1+f_2-f_i)}$ in \eqref{eq:ISRS_GNmodel}, and its spectral distribution, mathematically expressed as $\int G_{\text{Tx}}(\nu)e^{-P_{\text{tot}}C_{\text{r}} L_{\text{eff}}\nu} d\nu$ in \eqref{eq:ISRS_GNmodel}. 
\par 
\ 
In deriving \eqref{eq:XPM_integral} and \eqref{eq:SPM_contribution}, it was assumed that the total launch power of the entire WDM signal is uniformly distributed over the entire optical bandwidth. This effectively assumes that the effective channel power attenuation (the signal power profile) is \textit{independent} of the launch power distribution and only dependent on the total launch power. The impact of this assumption is discussed in more detail in the next section and quantified for mesh optical networks in Section \ref{sec:mesh_network} and sloped launch power distributions in Section \ref{sec:input_slope}.
\subsection{The ISRS GN model in closed-form}
\label{sec:closed-form}
The SPM and the XPM contribution are solved separately yielding two formulas, one for each contribution. The total NLI is then obtained using \eqref{eq:eta_total_XPM}. The reader is referred to Appendix \ref{XPM} and Appendix \ref{SPM} for the detailed derivations. 
\par  
\  
The closed-form approximation for the SPM contribution is
\begin{equation}
\begin{split}
&\eta_\text{SPM}\left(f_i\right) \approx \frac{16}{27}\frac{\gamma^2}{B^2_i} \left[ \frac{\pi\left(T_i^2 -\frac{4}{9}\right)}{ \alpha\phi_i}\text{asinh}\left(\frac{B_i^2\phi_i }{16\alpha}\right)+\frac{ B_i^2}{9\alpha^2}\right],
\label{eq:SPM_CF}
\end{split}
\end{equation}
with $\phi_i=12\pi^2\left(\beta_2+2\pi\beta_3f_i\right)$ and $T_i=2-\frac{f_iP_{\text{tot}}C_{\text{r}}}{\alpha}$. 
\par  
\  
The closed-form approximation for the total XPM contribution is
\begin{equation}
\begin{split}
&\eta_\text{XPM}\left(f_i\right) \approx \frac{32}{27}\frac{\gamma^2}{\alpha }\sum_{\forall k \in A_i} \left(\frac{P_k}{P_i}\right)^2\frac{1}{ B_k\phi_{i,k}} \\
&\left[\frac{T_k^2-1}{3}\text{atan}\left(\frac{B_i\phi_{i,k} }{\alpha}\right)+\frac{4-T_k^2}{6}\text{atan}\left(\frac{B_i\phi_{i,k}}{2\alpha}\right) \right]\\
\label{eq:XPM_CF}
\end{split}
\end{equation}
with $\phi_{i,k}=2\pi^2\left(f_k-f_i\right)\left[\beta_2+\pi\beta_3(f_i+f_k\right]$. The sum in \eqref{eq:XPM_CF} represents the summation over the XPM contribution of each individual interferer as in \eqref{eq:XPM_set}. Both formulas were derived from \eqref{eq:XPM_integral} with three key assumptions:
\par 
\ 
1) For the XPM contribution, the frequency separation between the channel of interest and the interfering channel is much greater than half of the channel bandwidth, i.e. $\left|f_k-f_i\right| = \left|\Delta f\right| \gg \frac{B_k}{2}$. 
\par 
2) The impact of ISRS on the effective channel power attenuation (i.e. the signal power profile) is small, which means that it can be approximated by a first-order Taylor series. 
\par 
3) The effective channel power attenuation (i.e. the signal power profile) is only a function of the total launch power and independent of its spectral distribution. This assumption has no impact on a uniform launch power distribution.
\par
\ 
\begin{figure}
\centering
\includegraphics[]{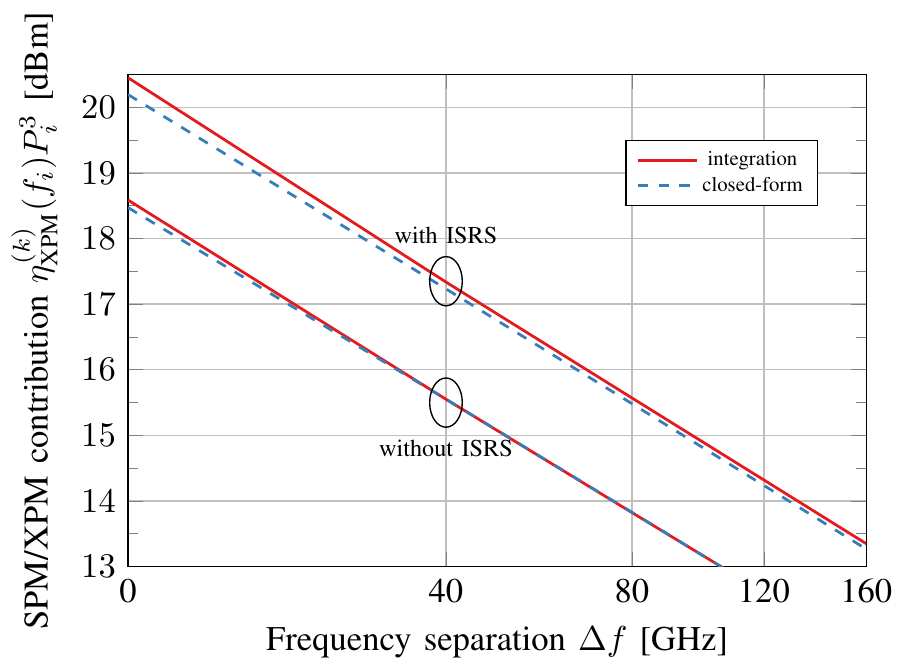}
\caption{XPM contribution in NLI power for channel $i=25$ with $f_i=-4000$ as a function of frequency separation between channel of interest (COI) and interferer (INF), obtained from numerically solving the ISRS GN model in integral form \eqref{eq:XPM_integral} and its proposed approximation in closed-form \eqref{eq:SPM_CF} and \eqref{eq:XPM_CF} (for $k\in\left\{26\text{,}\ 27\text{,}\ 28\text{,}\ 29\right\}$). A WDM signal with an optical bandwidth of 10~THz is assumed with $0$~dBm launch power per channel.}
\label{fig:NLI_contribution}
\end{figure}
Assumption 1) is valid when the interferer has a sufficient frequency separation from the COI. For very densely spaced channels, e.g. Nyquist-spacing, $\left|\Delta f\right| \gg \frac{B_k}{2}$ holds for all interferers that are separated by more than a few channel bandwidths. A separation of five channel bandwidths yields $5 \gg 0.5$. For non-Nyquist spacing, e.g. a ITU grid configuration of 32~GBd channels ($B_i \approx 32$~GHz) on a 50~GHz grid fulfills the same condition after only 3 channel separations. In reality, the impact of this assumption is even smaller (see Appendix \ref{XPM} for details). 
\par 
\  
\begin{figure*}[h]
\centering
\includegraphics[]{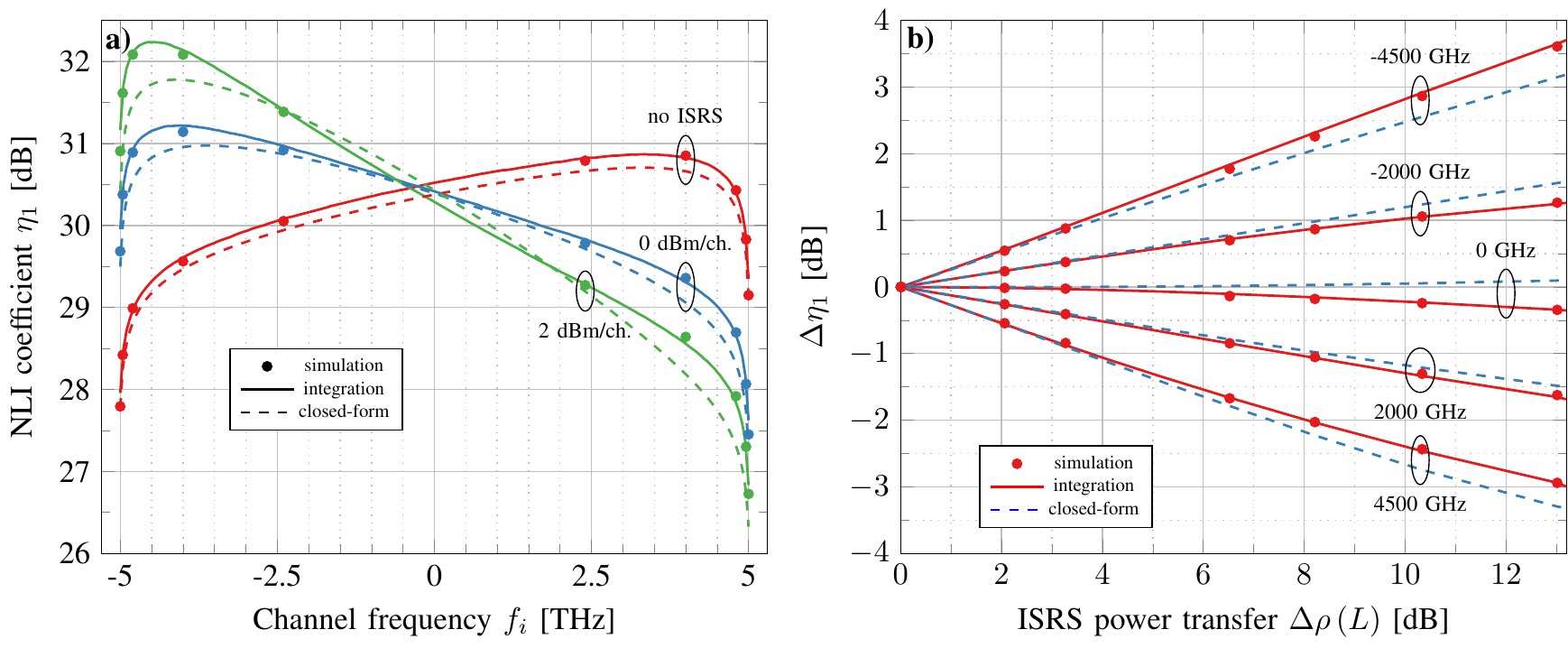}
\caption{The NLI coefficient after one span as a function of channel frequency is shown in a). The deviation of the NLI coefficient after one span as a function of ISRS power transfer for different channels within the transmitted WDM signal is shown in b). The results were obtained by numerical simulations, numerically solving the ISRS GN model in integral form \eqref{eq:ISRS_GNmodel} and its proposed approximation in closed-form \eqref{eq:SPM_CF} and \eqref{eq:XPM_CF}.}
\label{fig:eta_power1}
\end{figure*}
The SPM/XPM contribution as a function of channel separation, obtained by numerically integrating \eqref{eq:XPM_integral} and its proposed approximation in closed-form are shown in Fig. \ref{fig:NLI_contribution}. The COI has a center frequency of $f_i=-4000$~GHz and is transmitted with a single interferer over 100~km with parameters listed in Table \ref{tab:parameters}. Despite assumption 1), Eq. \eqref{eq:XPM_CF} yields remarkable accuracy with negligible error ($<0.1$~dB) for all XPM contributions with channel separations $\Delta f\geq 40$~GHz. The SPM contribution exhibits a discrepancy of only $0.1$~dB. However, the accuracy of the \textit{total} NLI is dominated by XPM which yields an accuracy of the proposed closed-form of $<0.1$~dB on the NLI (neglecting FWM) in the case of no ISRS. In the case of ISRS, the discrepancy is slightly higher due to assumption 2).
\par 
\
Assumption 2) holds when the impact of ISRS on the the effective channel power attenuation (i.e. the signal power profile) can be considered low. The SPM/XPM contribution for a COI with $f_i=-4000$~GHz and a launch power of $0$~dBm accounting for ISRS is shown in Fig. \ref{fig:NLI_contribution}. For this launch power, assuming a fully occupied 10 THz WDM signal, the power transfer between the outer channels is $\Delta\rho\left(L\right)\left[\text{dB}\right]=6.3$~dB and the net ISRS gain of the COI is $2.3$~dB. The only additional inaccuracy, compared to the case of no ISRS, originates from the weak ISRS assumption. The discrepancy for the SPM contribution is $0.3$~dB and $0.1$~dB for a frequency separation of $\Delta f=40$~GHz. The accuracy of this assumption depends on the strength of ISRS for a given transmission, where stronger ISRS reduces the accuracy. In order to obtain an approximate validity range, we compare the first-order and second-order terms of the Taylor expansion used to approximate ISRS. The reader is referred to Appendix \ref{limit} for a detailed derivation. It is found that the second-order term is negligible if and only if
\begin{equation}
\begin{split}
\label{eq:exact_ineq2}
\Delta \rho \left(L\right)\indB =4.3\cdot P_{\text{tot}}C_{\text{r}} L_{\text{eff}}B_{\text{tot}} \ll 26.
\end{split}
\end{equation}
For an ISRS power transfer of $\Delta \rho \left(L\right)\indB=6.3$~dB, Eq. \eqref{eq:exact_ineq2} yields $6.3 \ll 26$. For this case, \eqref{eq:exact_ineq2} is not fully satisfied and assumption 2) has a small impact on the accuracy of the closed-form expression.
\par 
Assumption 3) introduces no approximation error when the launch power distribution is uniform. When the launch power distribution deviates from a uniform one, the signal power profile is slightly changed due to ISRS. In practice, non-uniform launch power distributions may be present due to a variety of reasons. In mesh optical networks, non-continuous launch power distributions arise from unoccupied wavelengths as a result of traffic demands and routing algorithms. Additionally, launch power distributions might be optimized and sloped in order to achieve higher information throughput. The former case is addressed in Section \ref{sec:mesh_network} while the latter case is addressed in Section \ref{sec:input_slope}.
\section{Numerical Validation}
\label{sec:numerical_validation}
In this section the proposed closed-form approximations \eqref{eq:SPM_CF} and \eqref{eq:XPM_CF} are validated in an optical transmission system with parameters listed in Table \ref{tab:parameters}. The validation is performed for a point-to-point transmission in \ref{sec:p2p_links} and for a mesh optical network scenario in \ref{sec:mesh_network}.
\par 
\
The validation was carried out by numerically solving the Manakov equation using the well established split-step Fourier method (SSFM). Inter-channel stimulated Raman scattering was included in the SSFM by applying a frequency dependent loss at every linear step, so that the signal power profile altered by ISRS, is obtained. 
\par 
\    
A logarithmic step size distribution was implemented, where $0.25\cdot 10^6$ simulation steps were found to be sufficient for launch powers as high as $0$~dBm per channel and $1\cdot 10^6$ for launch powers as high as $3$~dBm/ch. Launch powers of up to $3$~dBm/ch. were considered in order to check the validity of the weak ISRS assumption (see assumption 2 in Section \ref{sec:closed-form}) for power transfers of up to $\Delta \rho \left(L\right)\indB=13$~dB. At the beginning of the fiber, the step size was as short as $21.5$~cm ($86$~cm), while the step size was $2.15$~m ($8.6$~m) for $3$~dBm/ch. ($0$~dBm/ch.) after 100~km of propagation. 
\par 
\  
Gaussian symbols, drawn from a circular-symmetric Gaussian distribution and uniform 64-QAM symbols were used for transmission. The former was chosen in order to verify the closed-form approximation while the latter was chosen to compare the performance to a standard modulation format. 
\par 
\  
The receiver consisted of digital dispersion compensation, ideal root-raised-cosine (RCC) matched filtering and constellation rotation. The SNR was ideally estimated as the ratio between the variance of the transmitted symbols $E[|X|^2]$ and the variance of the noise $\sigma^2$, where $\sigma^2=E[|X-Y|^2]$ and $Y$ represents the received symbols after digital signal processing. The nonlinear interference coefficient was then estimated via Eq. \eqref{eq:SNR}. In order to improve the simulation accuracy, four different data realizations were simulated and averaged for each transmission. 
\par 
\  
Ideal, noiseless amplifiers were considered to ease the NLI computation and for a fair comparison between numerical simulation and ISRS GN model.
\begin{table}
\centering
\caption{System Parameters}
\label{tab:parameters}
  \begin{tabular}{ l | c }
    \hline
   \textbf{Parameters} & \\  \hline
    Loss ($\alpha$) [dB/km]& 0.2 \\ \hline
    Dispersion ($D$) [ps/nm/km]& 17 \\ \hline
     Dispersion slope ($S$) [ps/$\text{nm}^2$/km]& 0.067 \\ \hline
    NL coefficient ($\gamma$) [1/W/km]& 1.2\\ \hline
    Raman gain slope ($C_{\text{r}}$) [1/W/km/THz]&  0.028 \\ \hline
    Raman gain ($C_{\text{r}}\cdot 14$ THz) [1/W/km]& 0.4 \\ \hline
    Symbol rate [GBd]&   40\\ \hline
    Channel Bandwidth ($B_{i}$) [GHz]&   $40.004$\\ \hline
    Channel Launch Power ($P_{i}$) [dBm]&   0\\ \hline
    Total Launch Power ($P_\text{tot}$) [dBm]&   24\\ \hline
    Channel spacing [GHz]&  40.005 \\ \hline
    Number of channels &  251 \\ \hline
    Optical bandwidth ($B_\text{tot}$) [THz]& $10.05$ \\ \hline
    Reference Wavelength [nm]& $1550$ \\ \hline
    Roll-off factor [\%]&  0.01  \\ \hline
    Number of symbols [$2^{x}$]& 17  \\ \hline
    Simulation steps per span 
       
       [$10^6$]& 0.25 to 1  \\ \hline
  \end{tabular}
\end{table}
\par 
\ 
\begin{figure*}[h!]
\centering
\includegraphics[]{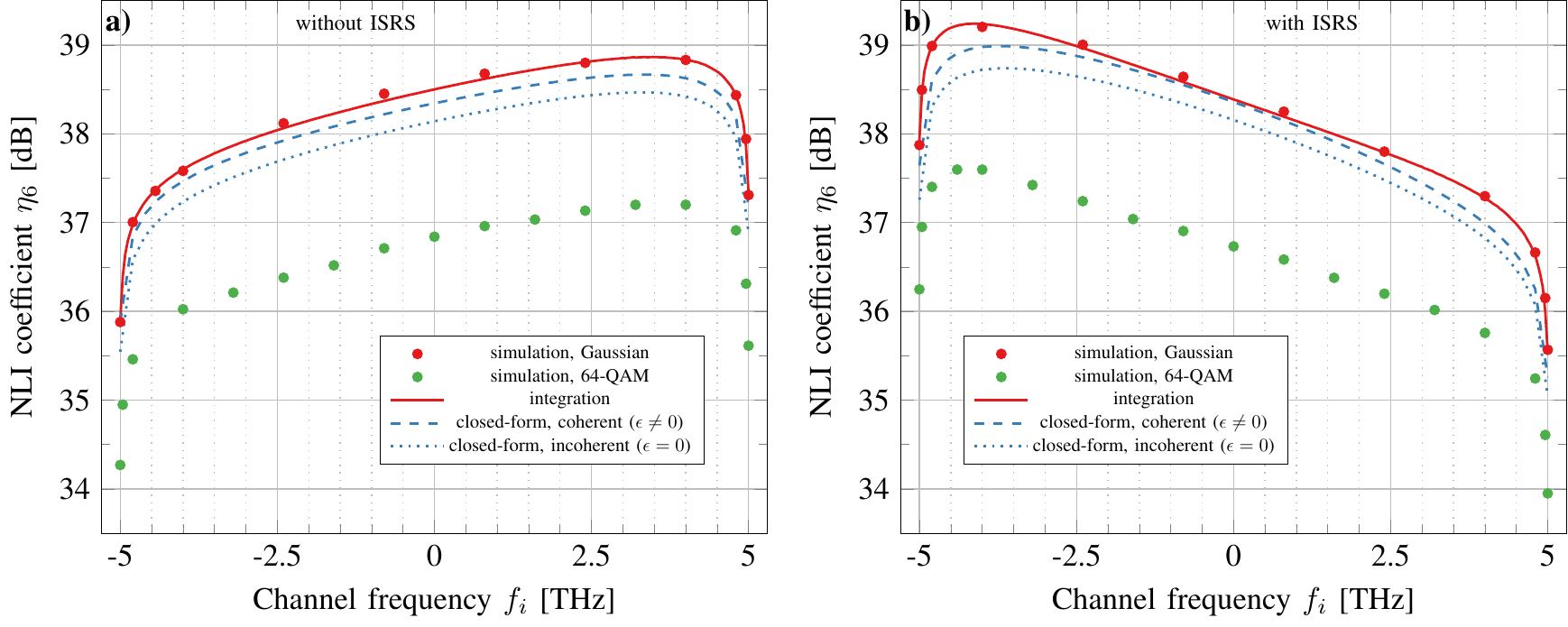}
\caption{The distribution of the NLI coefficient after 6 spans ($600$~km) without a) and with b) ISRS. A launch power of $0$~dBm/ch. was considered yielding an ISRS power transfer of $\Delta \rho \left(L\right)\indB=6.3$~dB. The results were obtained by numerical simulations, the ISRS GN model in integral form \eqref{eq:ISRS_GNmodel} and its proposed approximation in closed-form \eqref{eq:SPM_CF} and \eqref{eq:XPM_CF}, in coherent and incoherent form.}
\label{fig:eta_straight}
\end{figure*}
The spectral distribution of the NLI coefficient after a single span obtained by the SSFM, the ISRS GN model in integral form and its proposed approximation in closed form are shown in Fig. \ref{fig:eta_power1}a). Launch powers of $0$~dBm/ch. and $2$~dBm/ch. are shown which results in an ISRS power transfer of $\Delta \rho \left(L\right)\indB=6.3$~dB and $\Delta \rho \left(L\right)\indB=10.3$~dB, respectively. A case, where no ISRS is considered is shown for comparison. 
\par  
\  
The tilt in NLI, in the case of no ISRS, is due to the dispersion slope $S$ (or $\beta_3$), where low frequency components exhibit a higher amount of dispersion resulting in lower nonlinear penalties. With increasing launch powers, low frequency components are increasingly amplified, at the expense of high frequency components, leading to increased (reduced) NLI for low (high) frequency components. 
\par 
\ 
Not surprisingly, the ISRS GN model in integral form matches the simulation results with negligible error except at the most outer channels due to the local white noise assumption (which could be lifted by properly integrating the NLI PSD over the channel bandwidth). The proposed closed-form approximation is in good agreement with the ISRS GN model in integral form and the simulation results. The average gap, in the case of no ISRS, is $0.1$~dB. This discrepancy is due to the XPM assumption (see Section \ref{sec:xpm_assumption}), as the individual SPM and XPM contributions are approximated with negligible error, as shown in Fig. \ref{fig:NLI_contribution}. The average discrepancy is $0.1$~dB for $0$~dBm/ch. and $0.2$~dB for 2 dBm/ch. launch power. The increasing discrepancy with increasing launch power is due to the weak ISRS assumption (see Section \ref{sec:closed-form}). This assumption has more impact on the outer channels as the net ISRS gain is larger. 
Not surprisingly, the ISRS GN model in integral form matches the simulation results with negligible error except at the most outer channels due to the local white noise assumption (which could be lifted by properly integrating the NLI PSD over the channel bandwidth). The proposed closed-form approximation is in good agreement with the ISRS GN model in integral form and the simulation results. The average gap, in the case of no ISRS, is $0.1$~dB. This discrepancy is due to the XPM assumption (see Section \ref{sec:xpm_assumption}), as the individual SPM and XPM contributions are approximated with negligible error, as shown in Fig. \ref{fig:NLI_contribution}. The average discrepancy is $0.1$~dB for $0$~dBm/ch. and $0.2$~dB for 2 dBm/ch. launch power. The increasing discrepancy with increasing launch power is due to the weak ISRS assumption (see Section \ref{sec:closed-form}). This assumption has more impact on the outer channels as the net ISRS gain is larger for those channels. 
\par 
\ 
The deviation of the NLI coefficient as a function of the ISRS power transfer $\Delta \rho \left(L\right)\indB$ for different channels within the WDM signal is shown in Fig. \ref{fig:eta_power1}b). The discrepancy between the ISRS GN model in integral form and the SSFM is negligible for the shown range of power transfers. Due to the weak ISRS assumption, the accuracy of the closed-form expression decreases with increasing ISRS. This is because higher order terms of the Taylor expansion are becoming significant. 
\subsection{A point-to-point transmission scenario}
\label{sec:p2p_links}
In this section, a multi-span transmission system, consisting of six identical 100~km SMF fiber spans, is studied with parameters listed in Table \ref{tab:parameters}. A uniform launch power of $0$~dBm/ch. was used which is the optimum launch power for the central channel in the presence of Erbium-doped fiber amplifiers (EDFA) with a noise figure of 5~dB. The ISRS power transfer was equalized by a gain flattening filter after every fiber span.
\par 
\ 
The NLI coefficient after six spans is shown in Fig. \ref{fig:eta_straight}a) without accounting for ISRS and in Fig. \ref{fig:eta_straight}b) accounting for ISRS. Simulation results using Gaussian modulation as well as uniform 64-QAM are shown together with the ISRS GN model in integral form and its proposed approximation in closed-form. To account for coherent accumulation and variably loaded fiber spans, the ISRS GN model takes a slightly different form which was published in \cite[Eq. (2)]{Semrau_2018_tig} and used to obtain the results in Fig. \ref{fig:eta_straight}. 
\par 
\ 
The closed-form approximation is considered with an incoherent ($\epsilon=0$) and a coherent ($\epsilon \neq 0$) accumulation of NLI along multiple fiber spans (see \eqref{eq:eta_total_XPM}). The coherence factor of the given system configuration is $\epsilon=0.15$. The average gap between the closed-form, including a coherent accumulation, and the ISRS GN model in integral form is 0.1~dB and 0.2~dB without and with ISRS, respectively. The accuracy is similar to the single span case (see Fig. \ref{fig:eta_power1}), indicating that Eq. \eqref{eq:eta_total_XPM} sufficiently approximates the coherent accumulation of NLI. 
\par 
\    
A majority of the NLI originates from XPM, which is accumulating incoherently. The formalism may therefore be simplified by assuming an incoherent accumulation and setting $\epsilon = 0$. The average accuracy loss, of assuming incoherent accumulation, is $0.2$~dB for the studied system. Depending on accuracy requirements, this error may be deemed negligible. It should be noted, however, that this accuracy loss (with respect to Gaussian modulation) increases with the number of spans.   
\par 
\ 
A key assumption of the model is that each frequency component carries a symbol drawn from a symmetric circular Gaussian distribution which leads to an overestimation of the NLI power with respect to square QAM formats. To compare the model predictions to a standard modulation format, the NLI coefficient using 64-QAM obtained by the SSFM is shown in Fig. \ref{fig:eta_straight}. The average gap between SSFM using 64-QAM and the closed-form approximation in coherent form is $1.6$~dB in both cases, without and with ISRS. This gap decreases with increasing accumulated dispersion, hence, with increasing transmission distance. Additionally, modern transmission systems utilize probabilistic or geometric shaping which further decreases this gap as shaped signals partially resemble Gaussian modulated signals \cite{Forney_1984_emf,Fehenberger_2016_ops}.
\par 
\ 
The modulation format dependence can be partly accounted for with the following heuristic: First, it is assumed that ISRS has negligible impact on the \textit{relative} modulation format dependence, which is supported by the shown results. For lumped-amplified, multi-span transmission systems, where ISRS is negligible, the relative impact of the modulation format on the NLI has been derived in closed-form \cite[Eq. (8)]{Poggiolini_2015_asa}. Using this result, the average deviation can be reduced from $1.6$~dB to only $0.8$~dB. The modulation format dependence is not fully accounted for, as the formula is derived asymptotically for high span counts and it only corrects for XPM. Therefore, this approach is still conservative but with a reduced margin.
\par 
\
In summary, the proposed closed-form approximation models the impact of ISRS on the NLI with excellent accuracy in fully occupied point-to-point transmission scenarios. In can, therefore, be used for system design, optimization and real-time performance estimations of ultra-wideband transmission point-to-point links.
\subsection{A mesh optical network scenario}
\label{sec:mesh_network}
\begin{figure}[t]
\centering
\newcommand\W{0.5} 
\newcommand\WW{0.35} 
\newcommand\specheight{2.2} 
\newcommand\specwidth{4.6} 
\newcommand\nodecolor{black!10} 
\newcommand\ampcolor{white} 

\includegraphics[]{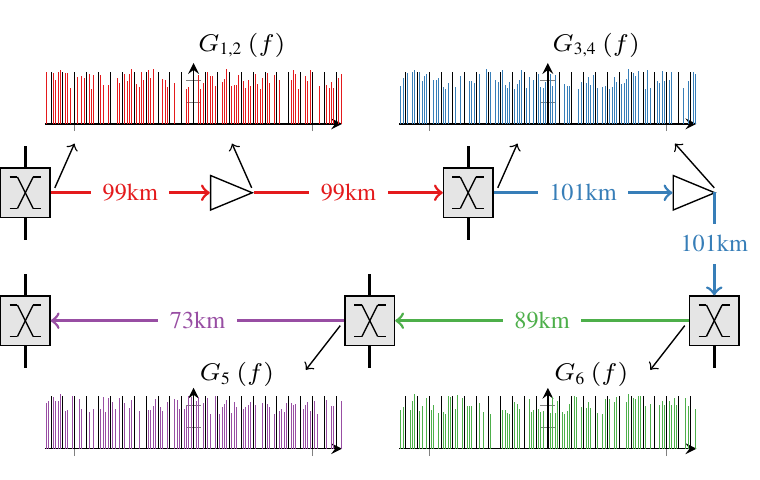}
\caption{Example transmission link, between nodes 15 and 13, from the BT 20+2 topology of the UK core network Fig. 6, showing interfering channels (in color) added and dropped at each ROADM. The channels of interest, that are transmitted along the entire lightpath, are shown in black.}
\label{fig:networkscheme}
\end{figure}

In this section, the closed-form approximation \eqref{eq:SPM_CF} and \eqref{eq:XPM_CF} is applied and validated in a mesh optical network. The fundamental difference in a mesh network, as opposed to a point-to-point transmission, is that not all channels within a WDM signal are transmitted along the \textit{entire} lightpath. At each reconfigurable optical add-drop multiplexer (ROADM), channels are added and dropped according to traffic demands and as the result of wavelength routing and lightpath assignment algorithms (RWA). 
\par 
\  
We introduce the following two definitions. For a given lightpath, channels that are transmitted along the entire lightpath are denoted as channels of interest. On the other hand, channels that are added and/or dropped at any point along the lightpath are denoted as interfering channels. Due to variably loaded network edges, the NLI of the channels of interest is different with respect to an equivalent point-to-point transmission as different interfering channels are emphasizing different XPM contributions $\eta_\text{XPM}^{\left(k\right)}(f_i)$ at each network edges. Additionally, most interfering channels have already propagated through part of the network, resulting in different amounts of accumulation dispersion compared to a point-to-point transmission.
\par 
\ 
The British Telecommunications 20+2 topology of the United Kingdom core network \cite{Ives_2014_qti} is considered for the analysis. For a given network topology and a given traffic demand, a vast number of feasible lightpath combinations are possible. For the sake of validation, only one lightpath is analyzed with two different network utilizations. We define network utilization as the average spectrum occupancy out of the entire available optical bandwidth. 
\par  
\  
The lightpath under test is the path between node A and B as indicated in Fig. \ref{fig:network_topology}. It is assumed that the first two edges, the first edge with $197$~km length and the second edge with $203$~km length, are each split into two fiber spans. The resulting lightpath of interest is illustrated in Fig. \ref{fig:networkscheme}, where after each ROADM a different spectrum is launch into the fiber due to the adding and dropping of interfering channels. The channels of interest, that are propagated along the entire lightpath, are shown in black where interfering channels are shown in color. 
\begin{figure*}
\centering
\includegraphics[]{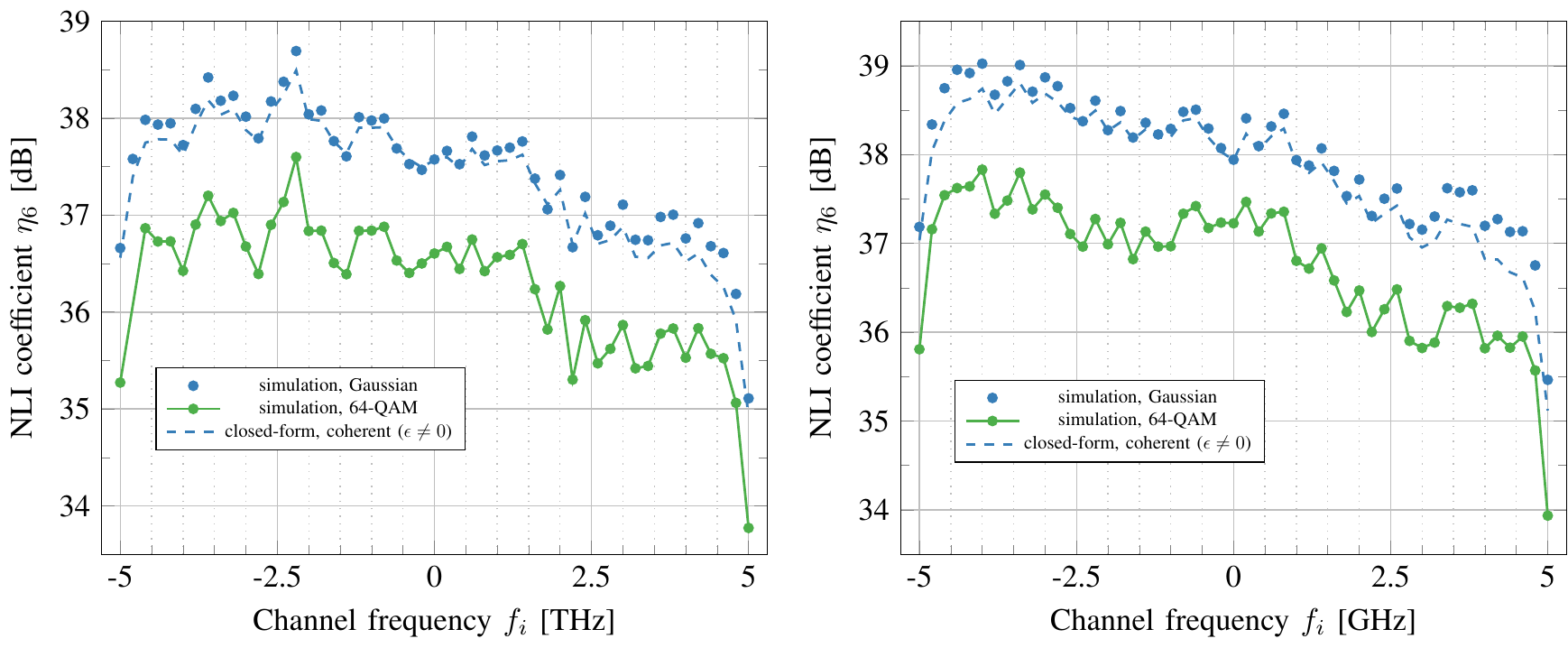}
\caption{The NLI coefficient of every fifth channel (i.e. a channel of interest) after six spans where interfering channels are continuously added and dropped along the transmission with a network utilization of 80\% shown in a) and 90\% shown in b). The results were obtained by numerical simulations and using the proposed closed-form approximation \eqref{eq:SPM_CF} and \eqref{eq:XPM_CF}.}
\label{fig:eta_network_80}
\end{figure*}
\par 
\ 
For this work, a few assumptions on the traffic and the established lightpaths are made. Every fifth channel (51 out of a total of 251 channels slots) was a channel of interest and their NLI coefficients were obtained by the SSFM and by the proposed closed-form approximation using \eqref{eq:SPM_CF} and \eqref{eq:XPM_CF}. The remaining 200 channel slots were partially filled with interfering channels which were continuously dropped and added at each ROADM. At each ROADM, 80\% of the interfering channels were randomly dropped and interfering channels were added by randomly choosing an empty channel slot. The unoccupied channel slots were randomly filled until a certain network utilization was reached. The considered network utilizations were 80\% and 90\%.
\ 
\par 
The added channels exhibit a random power offset between $\pm1$~dB  with respect to the COI to simulate potential non-ideal power equalization. Additionally, this was done in order to test the impact of assumption 3) (see Section \ref{sec:closed-form}) in a network scenario. Interfering channels were using the same modulation format as the channel under test and were randomly pre-dispersed corresponding to a transmission distance between 0 and 1000~km, to emulate the propagation from different lightpaths in the network. The wavelength dependent gain due to ISRS was ideally compensated after each span to ease a comparison to the point-to-point case in Section \ref{sec:p2p_links}. 
\par 
\ 
The NLI coefficient for a network utilization of $80$\% is shown in Fig. \ref{fig:eta_network_80}a) and a network utilization of $90$\% is shown in Fig. \ref{fig:eta_network_80}b). The SSFM results in Fig. \ref{fig:eta_network_80}a) were first published in \cite{Semrau_2018_tig}. The ISRS power transfers were $\Delta \rho \left(L\right)\indB=5$~dB and $\Delta \rho \left(L\right)\indB=5.7$~dB, which is less than in the point-to-point case as less average power was launched into a span.
\par 
\ 
The fluctuating behavior of the NLI coefficient is a direct consequence of the variably loaded network edges. The fluctuations are weaker in the case of $90$~\% network utilization as a larger average spectral occupation yields more averaging. The change in NLI due to ISRS was $-1.6$~dB to $1.5$~dB for 80\% of network utilization and $-1.8$~dB to $1.6$~dB for 90\% of network utilization. The proposed closed-form approximation is in good agreement with the simulation results with an average discrepancy of $0.1$~dB and $0.2$~dB for 80\% and 90\% network utilization, respectively. Assumption 3) in \ref{sec:closed-form} seems to have a negligible impact on the accuracy of the formula in variably loaded mesh optical networks. The average gap between the closed-form approximation and the SSFM using uniform 64-QAM is $1$~dB which is less than in the point-to-point case (cf. Fig. \ref{fig:eta_straight}) as interfering channels exhibit, in average, an higher amount of accumulated dispersion. 
\par 
\ 
Based on the validation carried out in this section, it is concluded that the proposed closed-form approximation models the NLI in mesh optical network scenarios with excellent accuracy. The results in this paper, therefore, enable the performance evaluation of complicated light path configurations for an entire network topology within only a few micro seconds. This is an essential step in the modeling of optical network performance in the ultra-wideband regime.
\section{Accuracy impact of sloped launch power distributions}
\label{sec:input_slope}
\begin{figure}
\centering
\includegraphics[]{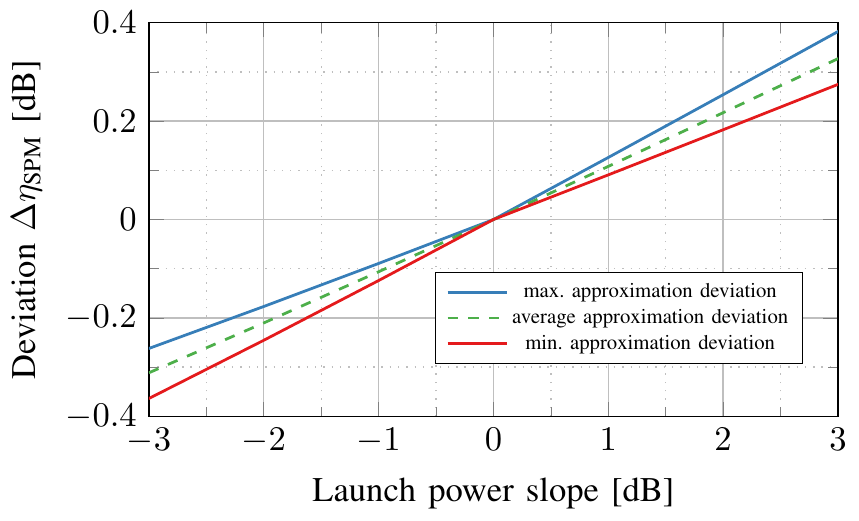}
\caption{Deviation between the SPM contribution by considering a uniform and a sloped launch power distribution as a function of input power slope for a 10~THz signal over a 100~km SMF span. The results were obtained using Eq. \eqref{eq:XPM_integral}.}
\label{fig:slope_approximation}
\end{figure}

In this section, the impact of a sloped launch power distribution on the accuracy of the proposed closed-form is addressed. In the derivation of the proposed closed-form approximation, it is assumed that the effective channel power attenuation is only a function of the total launch power and independent of its spectral distribution (see assumption 3 in Section \ref{sec:closed-form}). The accuracy loss in the case of variable loaded fiber spans was found to be negligible in Section \ref{sec:mesh_network}. 
\par 
\  
Sloped launch power distributions may be used as  in order to increase the achievable information throughput or to improve the SNR margin \cite{Cai_2015_4tt,Roberts_17_cpo,Galdino_18_1to}. 
\par 
\ 
The deviation of the SPM contribution using \eqref{eq:XPM_integral} as a function of launch power slope with respect to a uniform launch power distribution is shown in Fig. \ref{fig:slope_approximation}. A positive launch power slope means that high frequency channels have a larger launch power than low frequency channels. The plot shows maximum, minimum and average deviation of all channels within the WDM signal. Fig. \ref{fig:slope_approximation} can be interpreted as in the following. For a launch power slope of e.g. $\pm 2$~dB, assumption 3) in section \ref{sec:closed-form} introduces an approximation error of about $\mp 0.2$~dB on the predicted NLI.
\par 
\ 
A positively sloped launch power distribution increases the amount of measured NLI of all channels. This means that the gap between the NLI, predicted by the closed-form approximation, and the NLI of a square QAM signal actually \textit{decreases} as a function of input slope. This is due to a cancellation of approximation errors between assumption 3) in Section \ref{sec:closed-form} and the Gaussian modulation assumption. The opposite is true for negatively sloped launch power distributions.
\par 
\ 
The extension of the proposed closed-form approximation to fully account for arbitrary launch distributions is left for future research. 
\section{Conclusion}
A closed-form approximation of the Gaussian noise model in the presence of inter-channel stimulated Raman scattering was presented. 
\par 
\  
It was verified using split-step simulations and numerical integrations of the ISRS GN model in integral form, reporting an average deviation of 0.2~dB in nonlinear interference power for SMF based spans operating over the entire C+L band. This discrepancy, which could be addressed in the future, is primarily because the NLI contribution, that is jointly generated by two interfering channels (FWM or MCI), is neglected and due to the first-order description of ISRS. However, this has little impact for systems using dispersion unmanaged links, operating around optimum launch power.
\par 
\ 
The results in this paper allow for rapid evaluation of performance (e.g. SNR, maximum reach, optimum launch power) in ultra-wideband transmission systems, an essential step towards dynamic optical network capacity optimization and intelligent information infrastructure design.
\appendices
\section{Derivation of the XPM contribution}
\label{XPM}
In this section, the closed-form approximation of the XPM contribution \eqref{eq:XPM_CF} is derived. The derivation consists of finding an analytical approximation of the integral form \eqref{eq:XPM_integral} which models the nonlinear interference caused on channel $i$ by a single interfering channel $k$. The total XPM contribution is then obtained by summing over all XPM contributions as 
\begin{equation}
\begin{split}
\eta_\text{XPM}\left(f_i\right)=\sum_{\forall k}\eta_\text{XPM}^{\left(k\right)}\left(f_i\right).
\end{split}
\end{equation}
For notational brevity, we define $x\left(\zeta\right)=P_{\text{tot}}C_{\text{r}} L_{\text{eff}}\left(\zeta\right)$ and a pre-factor of $\frac{32}{27}\frac{\gamma^2}{B^2_k}\left(\frac{P_k}{P_i}\right)^2$ is suppressed throughout the derivation. Eq. \eqref{eq:XPM_integral} is then written as
\begin{equation}
\begin{split}
&\eta_\text{XPM}^{\left(k\right)}(f_i) = \int_{-\frac{B_i}{2}}^{\frac{B_i}{2}} df_1 \int_{-\frac{B_k}{2}}^{\frac{B_k}{2}} df_2 \ \Pi\left(\frac{f_1+f_2}{B_k}\right) \\
&\cdot \left| \int_0^L d\zeta \ \frac{xB_\text{tot}e^{-\alpha \zeta-x(f_1+f_2+f_i+\Delta f)}}{2\text{sinh}\left(\frac{xB_\text{tot}}{2}\right) }e^{j\phi\left(f_1+f_i,f_2+f_i+\Delta f,f_i,\zeta\right)}\right|^2\\
&\approx \int_{-\frac{B_i}{2}}^{\frac{B_i}{2}} df_1 \int_{-\frac{B_k}{2}}^{\frac{B_k}{2}} df_2 \ \left| \int_0^L d\zeta \ \frac{xB_\text{tot}e^{-\alpha \zeta-x(f_1+f_i+\Delta f)}}{2\text{sinh}\left(\frac{xB_\text{tot}}{2}\right) }\right.\\
&\cdot\left.e^{j\phi\left(f_1+f_i,f_i+\Delta f,f_i,\zeta\right)}\right|^2\\
&=2B_k\int_{0}^{\frac{B_i}{2}} df_1  \ \left| \int_0^L d\zeta \ \frac{xB_\text{tot}e^{-\alpha \zeta-x(f_1+f_i+\Delta f)}}{2\text{sinh}\left(\frac{xB_\text{tot}}{2}\right) }\right.\\
&\left.\cdot e^{j\phi\left(f_1+f_i,f_i+\Delta f,f_i,\zeta\right)}\right|^2,
\label{eq:xpm_1_step}
\end{split}
\end{equation}
where $\Delta f=f_k-f_i$ is the (center) frequency separation between channel $k$ and $i$. In \eqref{eq:xpm_1_step}, it is assumed that the frequency separation is much larger than half of the bandwidth of channel $k$ (i.e. $\left|\Delta f\right| \gg \frac{B_k}{2}$). This assumption allows to approximate $f_2 + \Delta f \approx \Delta f$. This approximation has only a small impact on the phase mismatch term $\phi$ for channels that are close to the COI. It has negligible impact on the ISRS term (the signal power profile) as it is essentially constant over one channel bandwidth $B_k$. Additionally, the approximation partly benefits from error cancellation, as the phase mismatch term is quasi affine with respect to $f_2$ over its integration domain $f_2\in \left[-\frac{B_k}{2}, \frac{B_k}{2}\right]$. In ultra-wideband transmission a large number of channels contribute to the total NLI, where the vast majority of interfering channels are spaced such that $\left|\Delta f\right| \gg \frac{B_k}{2}$ holds. Therefore the error of this assumption on the total NLI is expected to be small (see Fig. \ref{fig:NLI_contribution}). Additionally, the term $ g_k(f_1+f_2)$ in \eqref{eq:xpm_1_step} is neglected. 
\par 
The phase mismatch factor $\phi$, we obtain 
\begin{equation}
\begin{split}
&\phi\left(f_1+f_i,f_i+\Delta f,f_i,\zeta\right)\\
&=-4\pi^2f_1\Delta f\left[\beta_2+\pi\beta_3(f_1+2f_i+\Delta f)\right]\zeta\\
&\approx-4\pi^2f_1\Delta f\left[\beta_2+\pi\beta_3(2f_i+\Delta f)\right]\zeta\\
&=-4\pi^2f_1\left(f_k-f_i\right)\left[\beta_2+\pi\beta_3(f_i+f_k\right]\zeta\\
&=\phi_{i,k}f_1\zeta,
\end{split}
\end{equation}
with $\phi_{i,k}=-4\pi^2\left(f_k-f_i\right)\left[\beta_2+\pi\beta_3(f_i+f_k\right]$ and it is assumed that the impact of the dispersion slope is constant over one channel bandwidth $B_i$.
\par 
\ 
In order to simplify \eqref{eq:xpm_1_step}, the ISRS term is expanded into a Taylor series and truncated to first-order, assuming weak ISRS. The validity range of this approximation is analyzed in Appendix C in more detail and numerically validated in section \ref{sec:numerical_validation}. Additionally, it is assumed that the signal power profile is constant over one channel bandwidth $B_i$, mathematically $e^{-x(f_1+f_i+\Delta f)} \approx e^{-x(f_i+\Delta f)}$. The Taylor expansion of the ISRS term is then given by
\begin{equation}
\begin{split}
\frac{B_\text{tot}xe^{-x(f_i+\Delta f)}}{2\text{sinh}\left(\frac{B_\text{tot}}{2}x\right) } = 1 -(f_i+\Delta f)x+ \mathcal{O}(x^2),
\label{eq:taylor_series}
\end{split}
\end{equation}
and the signal power profile (to first-order) as
\begin{equation}
\begin{split}
&\frac{B_\text{tot}xe^{-\alpha\zeta-x(f_i+\Delta f)}}{2\text{sinh}\left(\frac{B_\text{tot}}{2}x\right) }\approx \left(1+T_k\right)e^{-\alpha \zeta} - T_ke^{-2\alpha \zeta},
\label{eq:ISRS_taylor}
\end{split}
\end{equation}
with
\begin{equation}
\begin{split}
T_k = -\frac{f_i+\Delta f}{\alpha}P_{\text{tot}}C_{\text{r}} = -\frac{P_{\text{tot}}C_{\text{r}}}{\alpha}f_k.
\end{split}
\end{equation}
Enabled by the first-order assumption of ISRS, the following simplification is obtained
\begin{equation}
\begin{split}
\label{eq:1}
&\left| \ \int_{0}^{L} d\zeta \ \frac{B_\text{tot}xe^{-\alpha\zeta-x(f_i+\Delta f)}}{2\text{sinh}\left(\frac{B_\text{tot}}{2}x\right) }e^{j\phi_{i,k}f_1\zeta}\right|^2 \\
&\approx\left| \ \int_{0}^{L} d\zeta \ \left(1+T_k\right)e^{-\alpha \zeta+j\phi_{i,k}f_1\zeta} - T_ke^{-2\alpha \zeta+j\phi_{i,k}f_1\zeta}\right|^2\\
&\approx \left| -\frac{\left(1+T_k\right)}{-\alpha +j\phi_{i,k}f_1} + \frac{T_k}{-2\alpha +j\phi_{i,k}f_1}\right|^2\\
&= \frac{(2+T_k)^2\alpha^2 }{4\alpha^4 +5\alpha^2\phi_{i,k}^2  f_1^2+\phi_{i,k}^4f_1^4 }+\frac{\phi_{i,k}^2f_1^2}{4\alpha^4 +5\alpha^2  \phi_{i,k}^2f_1^2+\phi_{i,k}^4f_1^4 },
\end{split}
\end{equation}
where it is assumed that $e^{-\alpha L}\ll 1$. Substituting the simplification \ref{eq:1} in \eqref{eq:xpm_1_step} and using the exact integral identities \eqref{eq:integral1} and \eqref{eq:integral2} yields
\begin{equation}
\begin{split}
&2B_k\int_{0}^{\frac{B_i}{2}} df_1  \ \left| \int_0^L d\zeta \ \frac{xB_\text{tot}e^{-\alpha \zeta-x(f_i+\Delta f)}}{2\text{sinh}\left(\frac{xB_\text{tot}}{2}\right) }e^{j\phi_{i,k}f_1\zeta}\right|^2\\
&\approx 2B_k \\
&\left[\frac{(2+T_k)^2-1}{3\alpha\phi_{i,k}}\text{atan}\left(\frac{\phi_{i,k} B_i}{2\alpha}\right)+\frac{4-(2+T_k)^2}{6\alpha\phi_{i,k}}\text{atan}\left(\frac{\phi_{i,k} B_i}{4\alpha}\right) \right].
\end{split}
\end{equation}
In order to obtain the XPM contribution of channel $k$ on channel $i$, the suppressed pre-factor $\frac{32}{27}\frac{\gamma^2}{B^2_k}\left(\frac{P_k}{P_i}\right)^2$ must be included. Finally, $T_k$ and $\phi_{i,k}$ are redefined and the individual XPM contributions $\eta_\text{XPM}^{\left(k\right)}(f_i)$ are summed up in order to obtain the total XPM contribution  $\eta_\text{XPM}(f_i)$ as in \eqref{eq:XPM_CF}.

\section{Derivation of the SPM contribution}
\label{SPM}
In this section, the closed-form SPM contribution of the NLI \eqref{eq:SPM_CF} is derived. The derivation consists of finding an analytical approximation of the integral expression \eqref{eq:XPM_integral} which models the nonlinear interference caused by channel $i$ on itself. The reader is reminded that, for the SPM contribution, a factor of $\frac{1}{2}$ must be multiplied to $\eqref{eq:XPM_integral}$. For notational brevity, we define $x\left(\zeta\right)=P_{\text{tot}}C_{\text{r}} L_{\text{eff}}\left(\zeta\right)$ and a pre-factor of $\frac{16}{27}\frac{\gamma^2}{B^2_i}$ is suppressed throughout the derivation.
\par 
The NLI coefficient of the SPM contribution is then written as
\begin{equation}
\begin{split}
&\eta_\text{SPM}(f_i) = \frac{1}{2}\eta^{\left(i\right)}_\text{XPM}(f_i) \approx \int_{-\frac{B_i}{2}}^{\frac{B_i}{2}} df_1\int_{-\frac{B_i}{2}}^{\frac{B_i}{2}} df_2\
\\
&\cdot\left| \ \int_{0}^{L} d\zeta \ \frac{xB_\text{tot}e^{-\alpha \zeta-xf_i}}{2\text{sinh}\left(\frac{xB_\text{tot}}{2}\right) }e^{j\phi\left(f_1+f_i,f_2+f_i,f_i,\zeta\right)}\right|^2  \\
&\approx 4 \int_{0}^{\frac{B_i}{2}} df_1\int_{0}^{\frac{B_i}{2}} df_2\
 \ \left| \ \int_{0}^{L} d\zeta \ \frac{xB_\text{tot}e^{-\alpha \zeta-xf_i}}{2\text{sinh}\left(\frac{xB_\text{tot}}{2}\right) }\right.\\
&\left.\cdot e^{j\phi\left(f_1+f_i,f_2+f_i,f_i,\zeta\right)}\right|^2,
\label{eq:SPM_first_step}
\end{split}
\end{equation}
where, again, it was assumed that the power transfer due to ISRS is constant over one channel bandwidth $B_i$.
\par 
\
For the phase mismatch factor $\phi$, we obtain
\begin{equation}
\begin{split}
&\phi\left(f_1+f_i,f_2+f_i,f_i,\zeta\right)\\
&=-4\pi^2f_1f_2\left[\beta_2+\pi\beta_3(f_1+f_2+2f_i)\right]\zeta\\
&\approx-4\pi^2f_1f_2\left(\beta_2+2\pi\beta_3f_i\right)\zeta\\
&=\phi_{i}f_1f_2\zeta,
\end{split}
\end{equation}
with $\phi_{i}=-4\pi^2\left(\beta_2+2\pi\beta_3f_i\right)$, where it is assumed that the impact of the dispersion slope is negligible over one channel bandwidth $B_i$.
\par 
\
Similar to the derivation of the XPM contribution, the ISRS term is expanded into a first-order Taylor series as in \eqref{eq:ISRS_taylor}. In order to derive an analytical approximation of the SPM contribution, it is necessary to further assume that $2\alpha^2 \gg \phi_i^2f_1^2f_2^2$. Eq. \eqref{eq:SPM_first_step} can then be approximated as  
\begin{equation}
\begin{split}
&4 \int_{0}^{\frac{B_i}{2}} df_1\int_{0}^{\frac{B_i}{2}} df_2\
 \ \left| \ \int_{0}^{L} d\zeta \ \frac{xB_\text{tot}e^{-\alpha \zeta-P_{\text{tot}}C_{\text{r}} L_{\text{eff}}f_i}}{2\text{sinh}\left(\frac{xB_\text{tot}}{2}\right) }\right.\\
&\left.\cdot e^{j\phi_if_1f_2\zeta}\right|^2\\
&\approx 4 \int_{0}^{\frac{B_\text{ch}}{2}} df_1\int_{0}^{\frac{B_\text{ch}}{2}} df_2\
 \ \frac{(2+T_i)^2\alpha^2 +\left(\phi_if_1f_2\right)^2}{\left[2\alpha^2-\left(\phi_if_1f_2\right)^2\right]^2 +9\alpha^2 \left(\phi_if_1f_2\right)^2 }\\
&\approx 4 \int_{0}^{\frac{B_\text{ch}}{2}} df_1\int_{0}^{\frac{B_\text{ch}}{2}} df_2\
 \ \frac{(2+T_i)^2\alpha^2 +\left(\phi_if_1f_2\right)^2}{4\alpha^4 +9\alpha^2 \left(\phi_if_1f_2\right)^2}
 \\
  &=4 \int_{0}^{\frac{B_i}{2}} df_1\left[\frac{\left(\frac{1}{6}(2+T_i)^2 -\frac{2}{27}\right)}{\phi_i \alpha}\frac{\text{atan}\left(\frac{3\phi_i}{4\alpha}B_if_1\right)}{f_1}+\frac{ B_i}{18\alpha^2}\right] \\
  &\approx 4 \left[ \frac{\left(\frac{1}{6}(2+T_i)^2 -\frac{2}{27}\right)}{\phi_i \alpha}\frac{\pi}{2} \text{asinh}\left(\frac{3\phi_i B_i^2}{16\alpha}\right)+\frac{ B_i^2}{36\alpha^2}\right].
\end{split}
\end{equation}
In the last derivation step, the exact integral identities \eqref{eq:integral3} and the approximate integral solution \eqref{eq:integral4} were used. Finally, the pre-factor $\frac{16}{27}\frac{\gamma^2}{B^2_i}$ must be included and $\phi_i$ is redefined in order to obtain the SPM contribution in closed-form as in \eqref{eq:SPM_CF}.

\section{Derivation of the validity range}
\label{limit}
In order to derive a validity range of the weak ISRS assumption, the ISRS term to first-order is compared to the ISRS term to second-order at a frequency component $f_k$. The first-order approximation is then valid when the second-order term is negligible.  
The second coefficient of the Taylor series, as in \eqref{eq:taylor_series}, is given by
\begin{equation}
\begin{split}
T^{(2)}_{k} &= \frac{f_k^2}{2}-\frac{B^2}{24}.
\end{split}
\end{equation}
Requiring that the second-order term is negligible to the first-order approximation yields
\begin{equation}
\begin{split}
\left|f_kx\right| &\gg \left|T^{(2)}_{k}x^2\right|=\left|\frac{f_k^2}{2}-\frac{B_\text{tot}^2}{24}\right|x^2.
\label{eq:validity_first_step}
\end{split}
\end{equation}
The channel that is most impacted by ISRS is the channel with center frequency $f_k=\frac{B}{2}$ for which we will evaluate \eqref{eq:validity_first_step} and obtain
\begin{equation}
\begin{split}
6 \gg B_\text{tot}P_\text{tot}L_\text{eff}C_r.
\label{eq:validity_second_step}
\end{split}
\end{equation}
Eq. \ref{eq:validity_second_step} can be related to the power transfer due to ISRS at the end of a fiber span $\Delta\rho\left(L\right)\left[\text{dB}\right]$ in decibels using \eqref{eq:max_pwrspread} as
\begin{equation}
\begin{split}
\label{eq:exact_ineq}
25.8 \gg \Delta\rho\left(L\right)\left[\text{dB}\right].
\end{split}
\end{equation}
It should be stressed that although $\Delta\rho\left(L\right)\left[\text{dB}\right]$ is expressed in decibels, the relation in \ref{eq:exact_ineq} compares two numerical numbers in a \textit{linear} manner. This means that for an ISRS power transfer of  $\Delta\rho\left(L\right)\left[\text{dB}\right]=6$ dB, we have $25.8 \gg 6$ \textit{not} $25.8\text{ dB} \gg 6\text{ dB}$.

\section{Integrals}
\label{integrals}
This section contains a lists of integral identities that were used in order to derive the proposed closed-form expression.
\begin{equation}
\begin{split}
\label{eq:integral1}
&\int_{0}^{X} dx \ \frac{1}{A +Bx^2+x^4 } \\
=&\frac{\sqrt[]{2}}{C\sqrt[]{B-C}}\text{atan}\left(\frac{\sqrt[]{2}X}{\sqrt[]{B-C}}\right)-\frac{\sqrt[]{2}}{C\sqrt[]{B+C}}\text{atan}\left(\frac{\sqrt[]{2}X}{\sqrt[]{B+C}}\right),\\
\end{split}
\end{equation}
\begin{equation}
\begin{split}
\label{eq:integral2}
&\int_{0}^{X} x \ \frac{x^2}{A +Bx^2+x^4 } \\
&=\frac{\sqrt[]{B+C}}{\sqrt[]{2}C}\ \text{atan}\left(\frac{\sqrt[]{2}X}{\sqrt[]{B+C}}\right)-\frac{\sqrt[]{B-C}}{\sqrt[]{2}C}\ \text{atan}\left(\frac{\sqrt[]{2}X}{\sqrt[]{B-C}}\right),\\
\end{split}
\end{equation}
with $C=\sqrt[]{B^2-4A}$.
\begin{equation}
\begin{split}
\label{eq:integral3}
\int_{0}^{X} df_2 \frac{1 +A^2f_2^2}{1 +B^2f_2^2} = \frac{\left(B^2-A^2\right)\text{atan}\left(Bx\right)+A^2BX}{B^3},
\end{split}
\end{equation}
\begin{equation}
\begin{split}
\label{eq:integral4}
&\int_{0}^{x} df_1 \ \frac{\text{atan}\left(Df_1\right)}{f_1}=\frac{1}{2}j\left[\text{Li}_2\left(-jDx\right)-\text{Li}_2\left(jDx\right)\right]\\
&\approx \frac{\pi}{2} \text{asinh}\left(\frac{Dx}{2}\right),
\end{split}
\end{equation}
where the approximation in \eqref{eq:integral4} was originally proposed in \cite{Poggiolini_2012_tgm}. A comparable asymptotic expansion of \eqref{eq:integral4} based on the natural logarithm was proposed in \cite[Appendix A]{Johannisson_2014_mon}.
\section*{Acknowledgment}
Financial support from UK EPSRC through programme grant TRANSNET (EP/R035342/1) and a Doctoral Training Partnership (DTP) studentship to Daniel Semrau is gratefully acknowledged.

\ifCLASSOPTIONcaptionsoff
  \newpage
\fi

\bibliographystyle{IEEEtran}
\bibliography{IEEEabrv,ref}

\begin{thebibliography}{10}
\providecommand{\url}[1]{#1}
\csname url@samestyle\endcsname
\providecommand{\newblock}{\relax}
\providecommand{\bibinfo}[2]{#2}
\providecommand{\BIBentrySTDinterwordspacing}{\spaceskip=0pt\relax}
\providecommand{\BIBentryALTinterwordstretchfactor}{4}
\providecommand{\BIBentryALTinterwordspacing}{\spaceskip=\fontdimen2\font plus
\BIBentryALTinterwordstretchfactor\fontdimen3\font minus
  \fontdimen4\font\relax}
\providecommand{\BIBforeignlanguage}[2]{{%
\expandafter\ifx\csname l@#1\endcsname\relax
\typeout{** WARNING: IEEEtran.bst: No hyphenation pattern has been}%
\typeout{** loaded for the language `#1'. Using the pattern for}%
\typeout{** the default language instead.}%
\else
\language=\csname l@#1\endcsname
\fi
#2}}
\providecommand{\BIBdecl}{\relax}
\BIBdecl

\bibitem{Hasegawa_2017_ofd}
T.~Hasegawa, Y.~Yamamoto, and M.~Hirano, ``Optimal fiber design for large
  capacity long haul coherent transmission,'' \emph{Opt. Express}, vol.~25,
  no.~2, pp. 706--712, Jan. 2017.

\bibitem{Semrau_2016_air}
D.~Semrau, T.~Xu, N.~A. Shevchenko, M.~Paskov, A.~Alvarado, R.~I. Killey, and
  P.~Bayvel, ``Achievable information rates estimates in optically amplified
  transmission systems using nonlinearity compensation and probabilistic
  shaping,'' \emph{Optics Lett.}, vol.~42, no.~1, p. 121, Dec. 2016.

\bibitem{Bosco_2011_aro}
G.~Bosco, P.~Poggiolini, A.~Carena, V.~Curri, and F.~Forghieri, ``Analytical
  results on channel capacity in uncompensated optical links with coherent
  detection,'' \emph{Opt. Express}, vol.~19, no.~26, pp. B440--B451, Dec. 2011.

\bibitem{Shevchenko_2016_air}
N.~A. Shevchenko, T.~Xu, D.~Semrau, G.~Saavedra, G.~Liga, M.~Paskov,
  L.~Galdino, A.~Alvarado, R.~I. Killey, and P.~Bayvel, ``Achievable
  information rates estimation for 100-nm {Raman}-amplified optical
  transmission system,'' in \emph{ECOC 2016; 42nd European Conference on
  Optical Communication}, Sept. 2016, pp. 1--3.

\bibitem{Anagnostopoulos_2007_pli}
V.~Anagnostopoulos, C.~T. Politi, C.~Matrakidis, and A.~Stavdas, ``Physical
  layer impairment aware wavelength routing algorithms based on analytically
  calculated constraints,'' \emph{Optics Communications}, vol. 270, no.~2, pp.
  247--254, Feb. 2007.

\bibitem{Nespola_2014_gvo}
A.~Nespola, S.~Straullu, A.~Carena, G.~Bosco, R.~Cigliutti, V.~Curri,
  P.~Poggiolini, M.~Hirano, Y.~Yamamoto, T.~Sasaki, J.~Bauwelinck, K.~Verheyen,
  and F.~Forghieri, ``{GN-Model} validation over seven fiber types in
  uncompensated {PM}-16{QAM} {Nyquist}-{WDM} links,'' \emph{IEEE Photon.
  Technol. Lett.}, vol.~26, no.~2, pp. 206--209, Jan. 2014.

\bibitem{Nespola_2015_evo}
A.~Nespola, M.~Huchard, G.~Bosco, A.~Carena, Y.~Jiang, P.~Poggiolini, and
  F.~Forghieri, ``Experimental validation of the {EGN}-model in uncompensated
  optical links,'' in \emph{Optical Fiber Communication Conference
  (OFC)}.\hskip 1em plus 0.5em minus 0.4em\relax Optical Society of America,
  2015, p. Th4D.2.

\bibitem{Galdino_2016_edo}
L.~Galdino, G.~Liga, G.~Saavedra, D.~Ives, R.~Maher, A.~Alvarado, S.~Savory,
  R.~Killey, and P.~Bayvel, ``Experimental demonstration of
  modulation-dependent nonlinear interference in optical fibre communication,''
  in \emph{ECOC 2016; 42nd European Conference on Optical Communication}, Sept
  2016, pp. 1--3.

\bibitem{Saavedra_2017_eio}
G.~Saavedra, M.~Tan, D.~J. Elson, L.~Galdino, D.~Semrau, M.~A. Iqbal, I.~D.
  Phillips, P.~Harper, N.~M. Suibhne, A.~D. Ellis, D.~Lavery, B.~C. Thomsen,
  R.~I. Killey, and P.~Bayvel, ``Experimental investigation of nonlinear signal
  distortions in ultra-wideband transmission systems,'' in \emph{2017 Optical
  Fiber Communications Conference and Exhibition (OFC)}, March 2017, pp. 1--3.

\bibitem{Saavedra_2017_eao}
G.~Saavedra, M.~Tan, D.~J. Elson, L.~Galdino, D.~Semrau, M.~A. Iqbal, I.~D.
  Phillips, P.~Harper, A.~Ellis, B.~C. Thomsen, D.~Lavery, R.~I. Killey, and
  P.~Bayvel, ``Experimental analysis of nonlinear impairments in fibre optic
  transmission systems up to 7.3 {THz},'' \emph{J. Lightw. Technol.}, vol.~35,
  no.~21, pp. 4809--4816, Nov. 2017.

\bibitem{Saavedra_2018_isr}
G.~Saavedra, D.~Semrau, M.~Tan, M.~A. Iqbal, D.~J. Elson, L.~Galdino,
  P.~Harper, R.~I. Killey, and P.~Bayvel, ``Inter-channel stimulated {R}aman
  scattering and its impact in wideband transmission systems,'' in
  \emph{Optical Fiber Communication Conference}.\hskip 1em plus 0.5em minus
  0.4em\relax Optical Society of America, 2018, p. Th1C.3.

\bibitem{Tang_2002_tcc}
J.~Tang, ``The channel capacity of a multispan {DWDM} system employing
  dispersive nonlinear optical fibers and an ideal coherent optical receiver,''
  \emph{J. Lightw. Technol.}, vol.~20, no.~7, p. 1095, Jul. 2002.

\bibitem{Poggiolini_2012_tgm}
P.~Poggiolini, ``The {GN} model of non-linear propagation in uncompensated
  coherent optical systems,'' \emph{J. Lightw. Technol.}, vol.~30, no.~24, pp.
  3857--3879, Dec. 2012.

\bibitem{Semrau_2018_tig}
D.~Semrau, E.~Sillekens, R.~I. Killey, and P.~Bayvel, ``The {ISRS} {GN} model,
  an efficient tool in modeling ultra-wideband transmission in point-to-point
  and network scenarios,'' in \emph{2018 European Conference on Optical
  Communication (ECOC), Tu4G.6}, Sep. 2018, pp. 1--3, pre--print available in
  arxiv:1808.00\,533.

\bibitem{Cantono_2018_oti}
M.~Cantono, D.~Pilori, A.~Ferrari, C.~Catanese, J.~Thouras, J.~L. Auge, and
  V.~Curri, ``On the interplay of nonlinear interference generation with
  stimulated {R}aman scattering for {QoT} estimation,'' \emph{J. Lightw.
  Technol.}, pp. 1--1, 2018.

\bibitem{Ives_2014_qti}
D.~J. Ives, A.~Lord, P.~Wright, and S.~J. Savory, ``Quantifying the impact of
  non-linear impairments on blocking load in elastic optical networks,'' in
  \emph{Optical Fiber Communication Conference}.\hskip 1em plus 0.5em minus
  0.4em\relax Optical Society of America, 2014, p. W2A.55.

\bibitem{Splett_1993_utc}
A.~Splett, C.~Kurtzke, and K.~Petermann, ``Ultimate transmission capacity of
  amplified optical fiber communication systems taking into account fiber
  nonlinearities,'' in \emph{1993 The European Conference on Optical
  Communication (ECOC)}, 1993.

\bibitem{Louchet_2003_amf}
H.~Louchet, A.~Hodzic, and K.~Petermann, ``Analytical model for the performance
  evaluation of {DWDM} transmission systems,'' \emph{IEEE Photonics Technology
  Letters}, vol.~15, no.~9, pp. 1219--1221, Sept 2003.

\bibitem{Chen_2010_cef}
X.~Chen and W.~Shieh, ``Closed-form expressions for nonlinear transmission
  performance of densely spaced coherent optical {OFDM} systems,'' \emph{Opt.
  Express}, vol.~18, no.~18, pp. 19\,039--19\,054, Aug. 2010.

\bibitem{Savory_2013_aft}
S.~J. Savory, ``Approximations for the nonlinear self-channel interference of
  channels with rectangular spectra,'' \emph{IEEE Photon. Technol. Lett.},
  vol.~25, no.~10, pp. 961--964, May 2013.

\bibitem{Johannisson_2014_mon}
P.~Johannisson and E.~Agrell, ``Modeling of nonlinear signal distortion in
  fiber-optic networks,'' \emph{J. Lightw. Technol.}, vol.~32, no.~23, pp.
  3942--3950, Dec 2014.

\bibitem{Poggiolini_2015_asa}
P.~Poggiolini, G.~Bosco, A.~Carena, V.~Curri, Y.~Jiang, and F.~Forghieri, ``A
  simple and effective closed-form {GN} model correction formula accounting for
  signal non-{Gaussian} distribution,'' \emph{J. Lightw. Technol.}, vol.~33,
  no.~2, pp. 459--473, Jan. 2015.

\bibitem{Semrau_2017_ace}
D.~Semrau, G.~Saavedra, D.~Lavery, R.~I. Killey, and P.~Bayvel, ``A closed-form
  expression to evaluate nonlinear interference in {R}aman-amplified links,''
  \emph{J. Lightw. Technol.}, vol.~35, no.~19, pp. 4316--4328, Oct 2017.

\bibitem{Semrau_17_ard}
D.~Semrau, R.~Killey, and P.~Bayvel, ``Achievable rate degradation of
  ultra-wideband coherent fiber communication systems due to stimulated {Raman}
  scattering,'' \emph{Opt. Express}, vol.~25, no.~12, pp. 13\,024--13\,034,
  Jun. 2017.

\bibitem{Semrau_2018_tgn}
D.~Semrau, R.~I. Killey, and P.~Bayvel, ``The {G}aussian {N}oise model in the
  presence of inter-channel stimulated {R}aman scattering,'' \emph{J. Lightw.
  Technol.}, vol.~36, no.~14, pp. 3046--3055, July 2018.

\bibitem{Roberts_17_cpo}
I.~Roberts, J.~M. Kahn, J.~Harley, and D.~W. Boertjes, ``Channel power
  optimization of {WDM} systems following {Gaussian} noise nonlinearity model
  in presence of stimulated {Raman} scattering,'' \emph{J. Lightw. Technol.},
  vol.~35, no.~23, pp. 5237--5249, Dec. 2017.

\bibitem{Cantono_2018_mti}
M.~Cantono, J.~L. Auge, and V.~Curri, ``Modelling the impact of {SRS} on {NLI}
  generation in commercial equipment: an experimental investigation,'' in
  \emph{Optical Fiber Communication Conference}.\hskip 1em plus 0.5em minus
  0.4em\relax Optical Society of America, 2018, p. M1D.2.

\bibitem{Zirngibl_1998_amo}
M.~Zirngibl, ``Analytical model of {R}aman gain effects in massive wavelength
  division multiplexed transmission systems,'' \emph{Electron. Lett.}, vol.~34,
  no.~8, pp. 789--790, Apr. 1998.

\bibitem{Mecozzi_2012_nsl}
A.~Mecozzi and R.-J. Essiambre, ``Nonlinear shannon limit in pseudolinear
  coherent systems,'' \emph{J. Lightw. Technol.}, vol.~30, no.~12, pp.
  2011--2024, Jun. 2012.

\bibitem{Secondini_2012_afc}
M.~Secondini and E.~Forestieri, ``Analytical fiber-optic channel model in the
  presence of cross-phase modulation,'' \emph{IEEE Photon. Technol. Lett.},
  vol.~24, no.~22, pp. 2016--2019, Nov. 2012.

\bibitem{Dar_2013_pon}
R.~Dar, M.~Feder, A.~Mecozzi, and M.~Shtaif, ``Properties of nonlinear noise in
  long, dispersion-uncompensated fiber links,'' \emph{Opt. Express}, vol.~21,
  no.~22, p. 25685, Oct. 2013.

\bibitem{Ives_2014_atp}
D.~J. Ives, P.~Bayvel, and S.~J. Savory, ``Adapting transmitter power and
  modulation format to improve optical network performance utilizing the
  gaussian noise model of nonlinear impairments,'' \emph{J. Lightw. Technol.},
  vol.~32, no.~21, pp. 4087--4096, Nov 2014.

\bibitem{Carena_2014_emo}
A.~Carena, G.~Bosco, V.~Curri, Y.~Jiang, P.~Poggiolini, and F.~Forghieri,
  ``{EGN} model of non-linear fiber propagation,'' \emph{Opt. Express},
  vol.~22, no.~13, p. 16335, Jun. 2014.

\bibitem{Zhang_2017_fae}
F.~Zhang, Q.~Zhuge, and D.~V. Plant, ``Fast analytical evaluation of fiber
  nonlinear noise variance in mesh optical networks,'' \emph{IEEE/OSA Journal
  of Optical Communications and Networking}, vol.~9, no.~4, pp. C88--C97, April
  2017.

\bibitem{Ives_2016_ati}
D.~J. Ives, A.~Alvarado, and S.~J. Savory, ``Adaptive transceivers in nonlinear
  flexible networks,'' in \emph{ECOC 2016; 42nd European Conference on Optical
  Communication}, Sept 2016, pp. 1--3.

\bibitem{Forney_1984_emf}
G.~Forney, R.~Gallager, G.~Lang, F.~Longstaff, and S.~Qureshi, ``Efficient
  modulation for band-limited channels,'' \emph{IEEE Journal on Selected Areas
  in Communications}, vol.~2, no.~5, pp. 632--647, September 1984.

\bibitem{Fehenberger_2016_ops}
T.~Fehenberger, A.~Alvarado, G.~B\"{o}cherer, and N.~Hanik, ``On probabilistic
  shaping of quadrature amplitude modulation for the nonlinear fiber channel,''
  \emph{J. Lightw. Technol.}, vol.~34, no.~21, pp. 5063--5073, Nov 2016.

\bibitem{Cai_2015_4tt}
J.~X. Cai, Y.~Sun, H.~Zhang, H.~G. Batshon, M.~V. Mazurczyk, O.~V. Sinkin,
  D.~G. Foursa, and A.~Pilipetskii, ``49.3 {Tb}/s transmission over 9100 km
  using {C}+{L EDFA} and 54 {Tb}/s transmission over 9150 km using
  hybrid-{Raman EDFA},'' \emph{J. Lightw. Technol.}, vol.~33, no.~13, pp.
  2724--2734, Jul. 2015.

\bibitem{Galdino_18_1to}
L.~Galdino, A.~Edwards, M.~Ionescu, J.~James, W.~Pelouch, E.~Sillekens,
  D.~Semrau, D.~Lavery, R.~I. Killey, S.~Barnes, P.~Bayvel, and S.~Desbruslais,
  ``120 {T}bit/s transmission over single mode fibre using continuous 91 nm
  hybrid {R}aman-{EDFA} amplification,'' \emph{arXiv:1804.01845}, 2018.

\end{thebibliography}

\begin{IEEEbiographynophoto}{Daniel Semrau}
(S’16) received the B.Sc. degree in electrical engineering from the Technical University of Berlin, Berlin, Germany, in 2013, the M.Sc. degree in photonic networks engineering from Scuola Superiore Sant’Anna, Pisa, Italy, and Aston University, Birmingham, U.K., in 2015. In 2015, he joined the Optical Networks Group, University College London, U.K., where he is currently working toward the Ph.D. degree. In 2018, Daniel was presented with the Graduate Student Fellowship award of the IEEE Photonics Society. His research interests are mainly focused on channel modeling, nonlinear compensation techniques, and ultra-wideband transmission for long-haul coherent optical communications.
\end{IEEEbiographynophoto}

\begin{IEEEbiographynophoto}{Robert I. Killey}
(SM’17) received the B.Eng. degree in electronic and communications engineering from the University of Bristol, Bristol, U.K., in 1992, the M.Sc. degree from University College London (UCL), London, U.K., in 1994, and the D.Phil. degree from the University of Oxford, Oxford, U.K., in 1998. He is currently a Reader in optical communications with the Optical Networks Group with UCL. His research interests include nonlinear fiber effects in WDM transmission, advanced modulation formats, and digital signal processing for optical communications. He has participated in many European projects, including ePhoton/ONe, Nobel, BONE and ASTRON, and national projects. He is currently a Principal Investigator in the EPSRC funded UNLOC project. He was with the technical program committees of many international conferences including European Conference on Optical Communication, Optical Fiber Communication Conference ACP, and OECC. He was an Associate Editor of the IEEE/OSA Journal of Optical Communications and Networking and is currently an Associate Editor of the Journal of Lightwave Technology.
\end{IEEEbiographynophoto}

\begin{IEEEbiographynophoto}{Polina Bayvel}
(F'10) received the B.Sc. (Eng.) and Ph.D. degrees in electronic and electrical engineering from University College London (UCL), London, U.K., in 1986 and 1990, respectively. Her Ph.D. research focused on nonlinear fiber optics and their applications. In 1990, she was with the Fiber Optics Laboratory, General Physics Institute, Moscow (Russian Academy of Sciences), under the Royal Society Postdoctoral Exchange Fellowship. She was a Principal Systems Engineer with STC Submarine Systems, Ltd., London, U.K., and Nortel Networks (Harlow, U.K., and Ottawa, ON, Canada), where she was involved in the design and planning of optical fiber transmission networks. During 1994–2004, she held a Royal Society University Research Fellowship with UCL, and, in 2002, she became a Chair in Optical Communications and Networks. She is currently the Head of the Optical Networks Group, UCL. She has authored or coauthored more than 290 refereed journal and conference papers. Her research interests include optical networks, high-speed optical transmission, and the study and mitigation of fiber nonlinearities.Prof. Payvel was the 2002 recipient of the Institute of Physics Paterson Prize and Medal for contributions to research on the fundamental aspects of nonlinear optics and their applications in optical communications systems. In 2007, she was the recipient of the Royal Society Wolfson Research Merit Award. He is a Fellow of the Royal Academy of Engineering (F.R.Eng.), the Optical Society of America, the U.K. Institute of Physics, and the Institute of Engineering and Technology. She is a Member of the Technical Program Committee (TPC) of a number of conferences, including Proceedings of European Conference on Optical Communication (ECOC) and Co-Chair of the TPC for ECOC 2005.
\end{IEEEbiographynophoto}

\end{document}